\documentclass[
preprint,
nofootinbib,
 amsmath,amssymb,
 aps,
prd,
]{revtex4-1}

\makeatletter
\renewcommand*\l@section{\@dottedtocline{1}{1.5em}{2em}}
\renewcommand*\l@subsection{\@dottedtocline{1}{4em}{1.5em}}
\renewcommand*\l@subsubsection{\@dottedtocline{1}{7.0em}{1.5em}}
\makeatother

\usepackage{lipsum}
\usepackage{dcolumn}
\usepackage{bm}
\usepackage{color}
\usepackage{graphicx,amsmath,amssymb}
\usepackage{tikz}
\usetikzlibrary{matrix}
\usepackage{slashed}

\newcommand{\eqnsplit}[1]{\begin{align}\begin{split}#1\end{split}\end{align}}
\newcommand{\eqn}[1]{\begin{align}#1\end{align}}
\newcommand{\eqns}[1]{\begin{align*}#1\end{align*}}

\definecolor{awesome}{rgb}{1.0, 0.13, 0.32}
\definecolor{applegreen}{rgb}{0.55, 0.71, 0.0}
\definecolor{darkpastelgreen}{rgb}{0.01, 0.75, 0.24}
\definecolor{azure(colorwheel)}{rgb}{0.0, 0.5, 1.0}
\definecolor{fluorescentyellow}{rgb}{0.8, 1.0, 0.0}
\definecolor{guppiegreen}{rgb}{0.0, 1.0, 0.5}
\definecolor{inchworm}{rgb}{0.7, 0.93, 0.36}
\definecolor{richelectricblue}{rgb}{0.03, 0.57, 0.82}
\definecolor{springgreen}{rgb}{0.0, 1.0, 0.5}
\definecolor{mediumcandyapplered}{rgb}{0.89, 0.02, 0.17}
\definecolor{scarlet}{rgb}{1.0, 0.13, 0.0}

\newcommand{\logb}[1]{\log \left( #1 \right)}
\newcommand{\expb}[1]{\exp \left( #1 \right)}

\newcommand{\bdelta}{\breve{\delta}}

\newcommand{\D}{\mathcal{D}}
\newcommand{\pD}{{\mathcal{D}}^p}
\newcommand{\psiR}{\psi_\mathrm{R}}
\newcommand{\psiT}{\psi_\mathrm{T}}
\newcommand{\twp}{\tilde \omega_+}
\newcommand{\twm}{\tilde \omega_-}

\newcommand{\lc}{\left(}
\newcommand{\rc}{\right)}

\usepackage{hyperref}
\hypersetup{
    bookmarks=true,         
    unicode=false,          
    pdftoolbar=true,        
    pdfmenubar=true,        
    pdffitwindow=false,     
    pdfstartview={FitH},    
    pdftitle={Weighting gates in circuit complexity and holography},    
    pdfauthor={I. Akal},     
    pdfcreator={I. Akal},   
    pdfnewwindow=true,      
    colorlinks=true,       
    linkcolor=blue,          
    citecolor=darkpastelgreen,        
    filecolor=blue,      
    urlcolor=blue           
}

\begin{document}
\title{Weighting gates in circuit complexity and holography}
\author{I. Akal}
\affiliation{II. Institute for Theoretical Physics\\ 
University of Hamburg\\
D-22761 Hamburg, Germany}

\email{ibrahim.akal@desy.de}
\date{\today}

\begin{abstract}
Motivated by recent studies of quantum computational complexity in quantum field theory and holography, 
we discuss how weighting certain classes of gates building up a quantum circuit more heavily than others does affect the complexity.
Utilizing Nielsen's geometric approach to circuit complexity, we investigate the effects 
for a regulated field theory for which the optimal circuit is a representation of $GL(N,\mathbb{R})$.
More precisely, we work out how a uniformly chosen weighting factor acting on the entangling gates affects the complexity and, particularly, its divergent behavior.
We show that assigning a higher cost to the entangling gates increases the complexity. 
Employing the penalized and the unpenalized complexities for the $\mathcal{F}_{\kappa=2}$ cost, we further find an interesting relation between the latter and the one based on the unpenalized $\mathcal{F}_{\kappa=1}$ cost.
In addition, we exhibit how imposing such penalties modifies the leading order UV divergence in the complexity. We show that appropriately tuning the gate weighting eliminates the additional logarithmic factor, thus, resulting in a simple power law scaling. We also compare the circuit complexity with holographic predictions, specifically, based on the complexity=action conjecture, and relate the weighting factor to certain bulk quantities.
Finally, we comment on certain expectations concerning the role of gate penalties in defining complexity in field theory and also speculate on 
possible implications for holography.
\end{abstract}

\begin{titlepage}
\maketitle
\thispagestyle{empty}
\end{titlepage}
\tableofcontents

\section{Introduction}
\label{sec:intro}
The anti de-Sitter (AdS)/conformal field theory (CFT) correspondence \cite{Maldacena:1997re} has substantially improved our understanding of strongly coupled systems and black hole (BH) micro states. As an explicit realization of the holographic principle \cite{tHooft:1993dmi,Susskind:1994vu}, it basically dictates how quantum gravitational physics can nonperturbatively be formulated within the language of a pure quantum field theory (QFT). 
However, despite the enormous progress since its first proposal, many aspects of the duality still remain deeply mysterious.

In recent years, new concepts from quantum information and quantum computation have further helped to advance our understanding of the mechanisms behind the AdS/CFT correspondence \cite{Rangamani:2016dms,Aaronson:2016vto,Almheiri:2014lwa}. The found relationships may be traced back to the much celebrated Bekenstein-Hawking entropy formula which relates an information theoretic quantity to a geometric characteristic of a BH; the area of the event horizon. Motivated by this striking prediction, one could already expect that a more fundamental information theoretic notion is related to other geometric properties of BHs. 

In fact, it has been shown that entanglement properties of the boundary CFT are directly related to certain geometric quantities in the bulk \cite{Ryu:2006bv,Ryu:2006ef,Hubeny:2007xt}. These findings have provided surprising evidence that quantum entanglement plays a profound role in the emergence of spacetime in a gravitational theory. Since then, understanding the relation between quantum entanglement and the emergence of semiclassical geometry is being actively worked on, see e.g. \cite{VanRaamsdonk:2010pw,Lashkari:2013koa,Faulkner:2013ica}.

In the context of BHs, such an information/geometry duality turns out to be even more astonishing. Namely, it has been shown that while the holographic entanglement entropy approaches a constant value with time during BH thermalization, certain bulk quantities such as the size of the Einstein-Rosen bridge (ERB) for the eternal AdS BH keeps increasing \cite{Hartman:2013qma}. In view of that conundrum, it has been proposed that the boundary quantity which continues evolving after thermal equilibrium is quantum computational complexity \cite{Susskind:2014moa}. 

Two separate conjectures have been proposed in the holographic context.
The first one, the complexity=volume (CV) proposal, states that complexity is proportional to the volume of a maximal codimension one surface in the bulk which extends to the boundary \cite{Susskind:2014rva,Stanford:2014jda,Roberts:2014isa}. A second, more precise one, is known as the complexity=action (CA) proposal \cite{Brown:2015lvg,Brown:2015bva}. It identifies the complexity of the boundary CFT state with the gravitational action evaluated on a specific bulk region known as the Wheeler-DeWitt (WDW) patch. The WDW patch is the domain of dependence of a bulk Cauchy surface attached to a specific time slice. 
Also note that the notion of complexity has recently been related to the spacetime volume of the WDW patch \cite{Couch:2016exn}.

\begin{figure}[h!]
  \centering
    \includegraphics[width=0.48\textwidth]{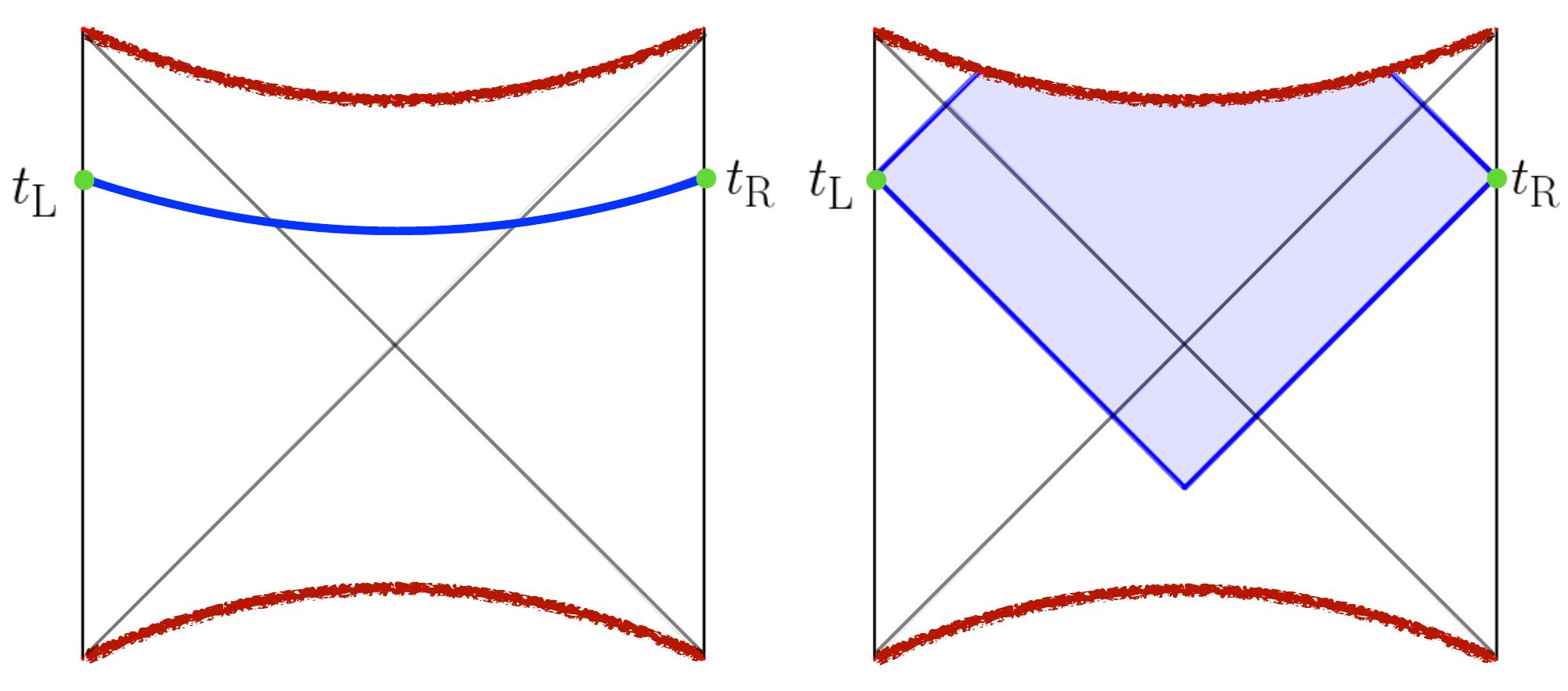}
      \caption{The extended Penrose diagrams illustrate the relevant geometric objects for the CV (left) and CA (right) proposals for the two sided eternal AdS BH which is dual to the thermofield double state (TFD) formed by entangling two copies of a CFT. In the left diagram the blue solid curve represents the maximal bulk hypersurface extending to the AdS boundary, asymptoting to the indicated $t_\text{L}$ and $t_\text{R}$ time slices on which the CFT state is defined. It connects the latter through the ERB. The shaded blue region in the right diagram represents the WDW patch where the CFT state is again evaluated on the mentioned time slices on the left and right boundaries.}
      \label{fig:cv-ca}
\end{figure}

A sketch of the relevant geometric objects for the CA and CV proposals is depicted in figure \ref{fig:cv-ca}. In both cases the corresponding quantity evolves with time even after thermal equilibrium sets in. Thus, allowing to probe the interior region of BHs, the mentioned proposals add two new classes of interesting gravitational observables with connections to quantum information and computation. 
Many aspects of the proposals and the related observables have recently been investigated, see e.g. \cite{Alishahiha:2015rta,Barbon:2015ria,Cai:2016xho,Brown:2016wib,Lehner:2016vdi,Yang:2016awy,Chapman:2016hwi,Carmi:2016wjl,Reynolds:2016rvl,Brown:2017jil,Zhao:2017iul,Alishahiha:2017hwg,Flory:2017ftd,Reynolds:2017lwq,Ghodrati:2017roz,Carmi:2017jqz,Couch:2017yil,Kim:2017qrq,Abt:2017pmf,Moosa:2017yvt,Moosa:2017yiz,Swingle:2017zcd,Reynolds:2017jfs,Fu:2018kcp,Chen:2018mcc,Chapman:2018dem,Abt:2018ywl,Hashimoto:2018bmb,Chapman:2018lsv,Flory:2018akz,Couch:2018phr,HosseiniMansoori:2018gdu,Goto:2018iay,Akhavan:2018wla,Chapman:2018bqj,Fan:2018xwf,Hashemi:2019xeq,Flory:2019kah,Yang:2019gce,Guo:2019vni,Bernamonti:2019zyy}. Even though both the CV and CA conjectures are interesting, much has to be done to advance our insights into the deep connection between quantum information and the structure of spacetime. For establishing the geometric dual of complexity,
a precise definition of this quantity in strongly coupled theories or even in more general QFTs is for sure necessary.

\begin{figure}[h!]
  \centering
    \includegraphics[width=0.48\textwidth]{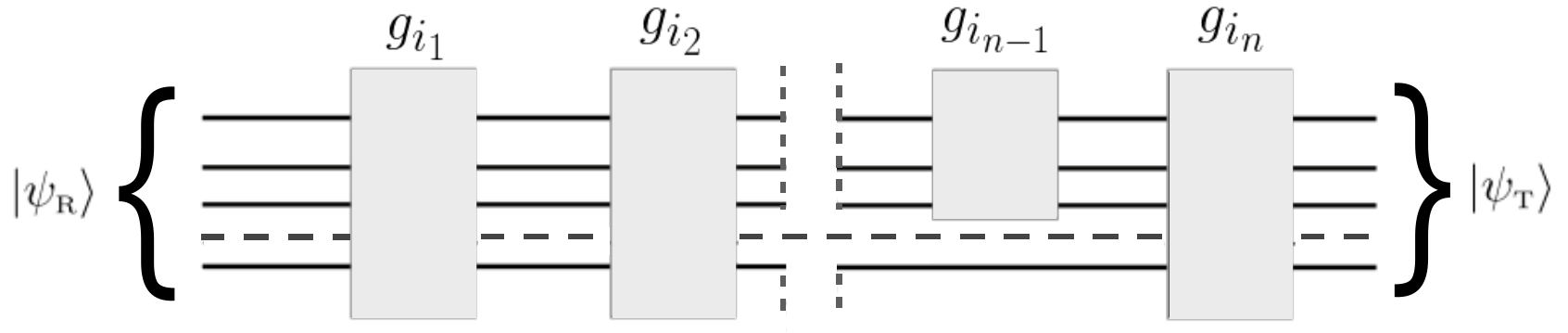}
      \caption{A quantum circuit constructing the target state $\vert \psi_\text{T} \rangle$ from the initial reference state $\vert \psi_\text{R} \rangle$. The unitary $U = g_{i_n} g_{i_{n-1}} \ldots g_{i_2} g_{i_1}$ is a sequence of elementary gates.}
      \label{fig:circuit}
\end{figure}

The concept of compuational complexity is rooted in the field of theoretical computer science. Generally, an implemented algorithm for mapping an input (reference) quantum state $\vert \psiR \rangle$ for a number of qubits to an output (target) quantum state $\vert \psiT \rangle$ is determined by some function which is a unitary operation $U$, i.e. $\vert \psiT \rangle = U \vert \psiR \rangle$. In a quantum circuit model, such a unitary operation (or circuit) is constructed from elementary gates selected from a fixed set of universal gates, see figure \ref{fig:circuit}. Accordingly, the circuit complexity can be defined as the minimal number of elementary gates required to construct the circuit. To define the circuit complexity for states in the boundary field theory \cite{Susskind:2014yaa,Brown:2017jil}, the corresponding task would then be identifying the minimum number of elementary gates required to prepare a desired target state by starting from a simpler reference state. For a construction built by a certain number of discrete elementary gates, the complexity of the target state explicitly depends on a specific reference, a choice of the gate set and an error tolerance. 

Preliminary progress towards quantifying circuit complexity in field theory has recently been made. For instance, the formulation in \cite{Jefferson:2017sdb} is based on earlier studies in quantum information showing that finding the optimal circuit is equivalent to finding the minimal geodesic in the space of unitaries \cite{nielsen2005geometric,nielsen2006quantum,dowling2008geometry}. This approach has also been generalized to fermionic \cite{Khan:2018rzm,Hackl:2018ptj} as well as interacting theories \cite{Bhattacharyya:2018bbv}; for other recent studies see e.g. \cite{Yang:2018nda,Alves:2018qfv,Magan:2018nmu,Guo:2018kzl,Chapman:2018hou,Ali:2018fcz,Ali:2018aon,Jiang:2018nzg,Sinamuli:2019utz,Liu:2019aji}. Another field theory proposal is based on the Fubini-Study metric approach \cite{Chapman:2017rqy} recently applied in e.g. \cite{Camargo:2018eof,Ali:2018fcz}. Even though both proposals have originally been made for free theories, it is remarkable that they give rise to similar leading order UV divergences as obtained in holographic computations. In addition, we also note that, motivated by the tensor network representation of the partition function, a notion of complexity has been proposed from the path integral point of view \cite{Caputa:2017urj,Caputa:2017yrh,Czech:2017ryf} which has given rise to interesting insights \cite{Takayanagi:2018pml}.

In the present paper, we focus on the notion of circuit complexity in QFT. More precisely, we investigate how weighting certain classes of gates building up a quantum circuit more heavily than others does affect the optimal circuit depth (i.e. complexity\footnote{We would like to note that the terms \textit{optimal circuit depth} and \textit{(circuit) complexity} are meant to mean the same. We use them interchangeably.}). 
Introducing such gate penalties may be motivated from different perspectives. Being already important concepts in quantum information and computation, such gate penalties may have interesting implications for complexity related aspects in QFT itself, but as well as for better understanding certain characteristics of quantum condensed matter systems.
Moreover, it may shed light on recent observations resulting from holographic approaches to complexity.
To be more precise, one may, for instance, ask for the consequence when complexity, if taken to be a physical attribute of a QFT, incorporates the notion of locality. In fact, if it is so, gates which entangle far separated points should require much higher costs in the geometric distance in the space of unitaries than those which operate as entanglers of less separated points. In the simple case, namely, for a pair of coupled harmonic oscillators, it has been shown that introducing such penalties has a drastic effect on the complexity \cite{Jefferson:2017sdb}. To be noted, there, the optimal circuit acts as a representation of $GL(2,\mathbb{R})$. 

Here, we work out how this change is reflected when similar penalties are introduced in the case of a field theory regulated by a lattice. In such a situation, the corresponding circuit becomes extended in form of a representation of $GL(N,\mathbb{R})$. We should note that instead of using distance dependent weighting factors, we work with entangling gates which are uniformly weighted. Of course, this does not entirely correspond to the situation described above, i.e. concerning the notion of locality. However, doing so generalizes previous findings for states in field theory. This will enable us to make interesting comparisons with previous findings. Beyond that, it has recently been argued that holographic complexity might be nonlocal, means that the gate set in the CFT contains bilocal gates acting at arbitraryly distant points \cite{Fu:2018kcp}. This additionally motivates us to penalize only one specific class of gates without implementing any distance dependencies.

Another motivation for considering the setup detailed above goes back to the found similarities between holographic complexity and circuit complexity in field theory.
Particularly, we would like to understand how introducing gate penalties, even if only acting uniformly, will affect the leading order UV divergence.
Such studies may be helpful for further comparisons with predictions from the gravity side. 
Closely related, introducing gate penalties may also be of particular importance for tensor network descriptions 
such as MERA \cite{Vidal:2008zz} or cMERA \cite{Haegeman:2011uy,Cotler:2018ufx}.
It has been argued that these constructions provide a representation of a time slice of the AdS space \cite{Swingle:2009bg,Swingle:2012wq}.

The paper is structured as follows:
in section \ref{sec:cc}, we start with a short description of the geometric approach to circuit complexity proposed by Nielsen and collaborators \cite{nielsen2005geometric,nielsen2006quantum,dowling2008geometry}. 
In section \ref{sec:setup}, we first introduce the setup under consideration and some basic notations. We then comment on the general procedure and summarize some of the earlier results. We would like to note that even though much of the content in this part has already been presented in \cite{Jefferson:2017sdb}, we recap the relevant findings which are needed for later computations. We also discuss some additional aspects which are important at the later stage. 
In section \ref{sec:pen-osc}, we go through previous findings concerning the effect of implementing gate penalties in the case of two coupled oscillators. 
Afterwards, we comment on some further details and extensions related to other cost functions.
Section \ref{sec:pen-qft} is the main part of this paper. We generalize the computations in penalized geometry for the regulated field theory and work out the differences to the unpenalized case. We then compare our results with holographic complexity computations, particularly, by referring to the findings based on the CA proposal. 
We also briefly comment on certain expectations concerning the implementation of distance dependent gates and on earlier findings which may share certain similarities with the presented results.
In section \ref{sec:conc}, we finalize with a brief conclusion.
In the remaining part, we mostly stick to the notations in \cite{Jefferson:2017sdb} to allow for a direct comparison with the unpenalized case. 

\section{Circuit complexity}
\label{sec:cc}
Let us begin by introducing the main idea of gemeotrizing circuit complexity {\`a} la Nielsen and collaborators.\footnote{For more details we refer the interested reader to \cite{Jefferson:2017sdb}.} The basic task is finding the optimal quantum circuit which implements a specific unitary operation $U$. This can be approached as a control problem of finding the right Hamiltonian  which constructs the desired unitary or circuit. For doing so, one specifies a space of unitaries where the interesting paths satisfy the boundary conditions $U(t=0) = \mathbf{1}$ and $U(t=1) \equiv U$.\footnote{The notation $\mathbf{1}$ stands for the identity matrix.} In addition, one also defines the corresponding cost function, call it $F$. Afterwards, by minimizing the following functional
\eqn{
\D(U(t)) = \int_0^1 dt\ F\left( U(t),\dot{U}(t) \right)
\label{eqn:D(U)}
}
one obtains the optimal path (or circuit depth) which yields the optimal quantum circuit. Here, the mentioned cost $F(U,V)$ is a local function of $U$ in the space of unitaries and $V$ is a vector in the tangent space at the point $U$. It is argued that $F$ must be (I) continuous, (II) positive, (III) homogeneous and (IV) should satisfy the triangle inequality. In the case when (I) is replaced with the criteria of smoothness, $\D(U)$ defines a length in a Finsler manifold which corresponds to a class of differential manifolds with a quasimetric and the measure as defined above. Thus, a Finsler geometry generalizes the notion of a Riemannian manifold. According to this instruction, the problem of finding the optimal circuit becomes the problem of finding the geodesics in the resulting geometry. 
What remains to be done is fixing the form of the cost $F$.
Once $F$ is chosen appropriately, the complexity is simply defined as the length of the optimal path in the corresponding geometry.

\section{Setup}
\label{sec:setup}
In the following, we specify the system we would like to focus on. We consider a free $d$ dimensional scalar field theory described by the Hamiltonian $H$ given by
\eqn{
H = \frac{1}{2} \int d^{d-1}x\ \left(  \pi^2(x)  + \vec{\nabla}\phi^2(x) + m^2 \phi^2(x)  \right).
}
We regulate this theory by putting it on a square lattice with spacing $\delta$ which describes an infinite number of coupled harmonic oscillators. The corresponding Hamiltonian then takes the form
\eqnsplit{
H &= \frac{1}{2} \sum_{\vec n} \left( \frac{P^2(\vec n)}{M} + M \omega^2 X^2(\vec n)
+ M \Omega^2    \sum_j \left( X(\vec n) - X(\vec n - \hat x_j) \right)^2 \right)
}
where the introduced definitions in $H$ above read as follows
\eqnsplit{
X(\vec n) = \delta^{d/2} \phi(\vec n),\quad P(\vec n) = \delta^{-d/2} p(\vec n),\\
 M=1/\delta,\quad \omega = m,\quad \Omega = 1/\delta.
 \label{eqn:qft-parameters}
}
Let us first proceed with the simplest case, i.e. a system of two coupled oscillators.

\subsection{Pair of oscillators}
\label{subsec:2osc}

\subsubsection{Position space}
\label{subsubsec:pos-space}
The system of two coupled harmonic oscillators is described by the following Hamiltonian
\eqn{
H = \frac{1}{2} \left(  p_1^2 + p_2^2 + \omega^2 (x_1^2 + x_2^2  ) + \Omega^2 (x_1 - x_2)^2  \right)
\label{eqn:H-2osc}
}
which is written in terms of the physical variables
where $x_j$ denote the spatial positions and $M_j=1$ is set for simplicity. The eigenstates and eigenenergies of \eqref{eqn:H-2osc} can be solved by rewriting the problem in terms of two decoupled oscillators via changing from the original position basis to the normal mode basis expressed in terms of  the following normal mode coordinates and frequencies
\eqnsplit{
\tilde x_\pm &:= \frac{1}{\sqrt{2}} (x_1 \pm x_2),\quad
\tilde \omega^2_+ := \omega^2,\quad\\
\tilde \omega_-^2 &:= \omega^2 + 2 \Omega^2.
\label{eqn:nm-variables}
}
The normalized\footnote{The normalization is chosen such that $\int d^2x\ |\psi_0|^2 = 1$.} ground state wave function can be written as the product of the ground state wave functions for the two individual oscillators. In terms of the physical position coordinates, the resulting wave function reads
\eqn{
\psi_0(x_1,x_2) = \frac{(\omega^2 - \beta^2)^{1/4}}{\sqrt{\pi}} e^{-\frac{\omega}{2}(x_1^2 + x_2^2) - \beta x_1 x_2}
}
where 
\eqnsplit{
\omega = \frac{\tilde \omega_+ + \tilde \omega_-}{2},\quad
\beta = \frac{\tilde \omega_+ - \tilde \omega_-}{2}.
}
Following the same motivation as described in \cite{Jefferson:2017sdb}, we choose a factorized Gaussian reference state where both oscillators are disentangled, i.e.
\eqn{
\psiR(x_1,x_2) = \frac{\sqrt{\omega_0}}{\sqrt{\pi}} e^{-\frac{\omega_0}{2} ( x_1^2 + x_2^2)  }.
}
Here, $\omega_0$ denotes some arbitrary frequency characterizing the reference state. Its explicit form will be discussed later. In order to construct $U$ such that 
\eqn{
\vert \psi_0 \rangle \equiv \vert \psiT \rangle = U \vert \psiR \rangle,
\label{eqn:R=UR}
}
we need to fix the set of appropriate gates. A set of elementary gates which implement a unitary transformation as prescribed in \eqref{eqn:R=UR} can be found in the literature. Without going into the details, here we will only mention that for our purpose the relevant one are the entangling and scaling gates, i.e.
\eqnsplit{
Q_{ab} &= e^{i \epsilon x_a p_b}\ \forall\ a \neq b\qquad (\text{entangling}),\\
Q_{aa} &= e^{\epsilon / 2} e^{i \epsilon x_a p_a}\qquad (\text{scaling}),
\label{eqn:gates}
}
noting that $a,b \in \{1,2\}$. The parameter $\epsilon$ appearing in \eqref{eqn:gates} is chosen to be infinitesimal\footnote{Note that due to  $\epsilon \ll 1$ we only need to find the circuit which minimizes the coefficient of the leading $1/\epsilon$ term in the complexity.} in order ensure only small changes when the gates act on the wave function.

Next, we may express the circuit $U$ in path ordered form,
\eqn{
U(s) = \overset{\leftarrow}{\mathcal P} \exp \left(   \int_0^s d\tilde s\ Y^I (\tilde s) \mathcal{O}_I \right).
\label{eqn:path-ordered-OP}
}
This representation will be used instead of a discrete gate representation as considered in the original formulation. The operators $\mathcal{O}_I$ precisely correspond to the previous gates from \eqref{eqn:gates} with $Q_{ab} = \exp \left( \epsilon \mathcal{O}_{ab}  \right)$ and $\mathcal{O}_{ab} = (i x_a p_b +\delta_{ab}/2)$ where $ab \equiv I \in \{ 11,12,21,22 \}$. Accordingly, the path ordered\footnote{The operators at smaller $\tilde s$ act prior to those at larger $\tilde s$.} exponential \eqref{eqn:path-ordered-OP} parametrizes the product of gates where the function $Y^I$ decides whether the gate of type $I$ is switched on or off. Note that the differential $d\tilde s$ behaves analogous to the infinitesimal parameter $\epsilon$. By proceeding in this way, any circuit follows a particular trajectory determined by $Y^I$ through the space of unitaries such that $ \psiT = U(s=1) \psiR$ and $U(s=0) = \mathbf{1}$. In the following, we choose the cost $F$ to be set by the following two norm
\eqn{
F = \mathcal F_2 \equiv \sqrt{\sum_I (Y^I)^2}.
\label{eqn:F2}
}
For this choice, note that the optimal path is a geodesic in usual Riemannian geometry. So the functional for the circuit depth from \eqref{eqn:D(U)} can be written as 
\eqn{
\mathcal{D} (U) = \int_0^1 d\tilde s\ \sqrt{G_{IJ} Y^I(\tilde s) Y^J(\tilde s)}
\label{eqn:circ-depth}
}
where we assume that the metric\footnote{In general, the metric $G_{IJ}$ allows to weight particular gates in the circuit. This will be discussed in the sections below.} is fixed by setting $G_{IJ} = \delta_{IJ}$.
Now, one can in principle express the function $Y^I$ which specifies the velocity vector tangent to the optimal trajectory in the space of unitaries in terms of the operators $\mathcal{O}_I$. However, it turns out to be more convenient to work with unitary matrices instead of operators. Accordingly, one just needs to reformulate the problem in matrix representation. In case of Gaussian states, as it is the case for the present setup, one may think of the space of states as the space of positive quadratic forms, i.e. $\psi \sim \exp \left( - \frac{1}{2} x_a A_{ab} x_b \right)$ where 
\eqn{
A_\mathrm{R} = \omega_0 \mathbf{1},\quad A_\mathrm{T} = \begin{pmatrix}
\omega & \beta \\ 
\beta & \omega
\end{pmatrix}.
}
It can be shown that the corresponding gate matrices take the form
\eqn{
Q_{I} = \exp \left(  \epsilon M_{I} \right)
}
where $M_{ab,cd} = \delta_{ac}\delta_{bd}$ denotes a $2 \times 2$ matrix. For the present setup, the explicit generators $M_I$ are 
\eqnsplit{
M_{11} &= \begin{bmatrix}
1 & 0 \\ 
0 & 0
\end{bmatrix},\quad
M_{22} = \begin{bmatrix}
0 & 0 \\ 
0 & 1
\end{bmatrix},\\
M_{12} &= \begin{bmatrix}
0 & 1 \\ 
0 & 0
\end{bmatrix} ,\quad
M_{21} = \begin{bmatrix}
0 & 0 \\ 
1 & 0
\end{bmatrix}.
\label{eqn:M_ij}
}
In terms of the gates from \eqref{eqn:M_ij}, it becomes clear that the circuits form a representation of $GL(2, \mathbb{R})$. Using the generator matrices in \eqref{eqn:M_ij}, the path ordered exponential in \eqref{eqn:path-ordered-OP} can be replaced by
\eqn{
U(s) = \overset{\leftarrow}{\mathcal{P}}  \int_0^s d\tilde s\ Y^I(\tilde s) M_I
\label{eqn:path-ordered-Mat}
}
where 
\eqn{
Y^I(s) = \mathrm{tr}(\partial_s U(s) U^{-1} (s) M_I^T).
\label{eqn:Y_I-Mat}
}
The expression in \eqref{eqn:Y_I-Mat} drastically simplifies the form of the velocity vector $Y^I$ which determines the optimal circuit depth. However, in order use the expression from \eqref{eqn:Y_I-Mat}, it is required to construct an explicit parameterization of the mentioned transformations. After having done this, the main task reduces to finding the minimal\footnote{Note that by definition the complexity is determined by the shortest geodesic which yields the desired transformation. In general, one may find a family of geodesics with various lengths.} geodesic in the resulting metric on $GL(2, \mathbb{R})$ which connects the matrices $A_\mathrm{T}$ and $A_\mathrm{R}$ according to $A_\mathrm{T} = U(s=1) A_\mathrm{R} U^T(s=1)$.

A general parameterization can be obtained by constructing $U$ according to the decomposition $GL(2, \mathbb{R}) = \mathbb{R} \times SL(2, \mathbb{R})$ where certain coordinates, for instance, labeled by $\tau,\theta$ and $\rho$, are chosen on the subgroup $SL(2, \mathbb{R})$. A fourth coordinate, call it $y$, will be responsible for the remaining $\mathbb{R}$ fibre. In this way, inserting $U \equiv U(\tau,\theta,\rho,y)$ into
\eqn{
ds^2 = G_{IJ}\  \mathrm{tr}(dU U^{-1} M_I^T)\  \mathrm{tr}(dU U^{-1} M_J^T)
\label{eqn:ds2}
}
yields a right-invariant metric. To obtain the geodesic, one first identifies the corresponding Killing vectors. Afterwards, these can be used to find all conserved charges which simplify solving the underlying geodesic equations. By proceeding as described, it has been found that the shortest geodesic is given by \cite{Jefferson:2017sdb}
\eqn{
y = y_1 s,\quad
\rho = \rho_1 s,\quad
\theta = \tau = 0
\label{eqn:SL-geo}
}
where 
\eqnsplit{
y_1 &= \frac{1}{2} \log \left( \frac{\sqrt{\omega^2 - \beta^2}}{\omega_0}   \right),\\
\rho_1 &= \frac{1}{2} \mathrm{cosh}^{-1} \left( \frac{\omega}{\sqrt{\omega^2 - \beta^2}}   \right).
\label{SL-boundary}
}
Substituting this straight line geodesic---since the metric is flat---into the general expression for $U \equiv U(\tau,\theta,\rho,y) \in GL(2,\mathbb{R})$ leads to the following optimal (straight line) circuit 
\eqn{
U_0(s) = \exp \left( \begin{bmatrix}
y_1 & -\rho_1 \\ 
-\rho_1 & y_1
\end{bmatrix} 
s
\right),\qquad 0 \leq s \leq 1.
\label{eqn:U0}
}
Here, we have expressed $U_0$ from \eqref{eqn:U0} in exponential form which is more convenient for later purpose. For instance, using \eqref{eqn:U0}, we can immediately identify the corresponding velocity vector components, i.e. $Y^I$'s. Having determined the latter, we just need to insert them into \eqref{eqn:circ-depth} which finally yields the complexity
\eqn{
\mathcal{C}_2 \equiv \mathcal D_2(U_0) = \sqrt{2 y_1^2 + 2 \rho_1^2}
\label{eqn:C2}
}
for the $\mathcal F_2$ cost from \eqref{eqn:F2} as we have indicated by the subscript.

\subsubsection{Normal mode subspace}
\label{subsubsec:nm-space}
Of course, one can express the complexity $\mathcal{C}_2$ from \eqref{eqn:C2} in terms of the physical parameters. For this, we just need to use the relations in \eqref{SL-boundary}. However, the final result would look rather complicated. Instead, in terms of the normal mode frequencies introduced in \eqref{eqn:nm-variables}, which can be used to express \eqref{SL-boundary} as 
\eqn{
y_1 = \frac{1}{4} \log \left( \frac{\tilde \omega_+ \tilde \omega_-}{\omega_0^2} \right),\quad \rho_1 = \frac{1}{4} \log \left( \frac{\tilde \omega_-}{\tilde \omega_+} \right),
\label{eqn:y1-rho1-nm}
}
the complexity in \eqref{eqn:C2} takes the form
\eqn{
\mathcal{C}_2 = \mathcal D_2(U_0) = \frac{1}{2} \sqrt{ \log^2\left( \frac{\tilde \omega_+}{\omega_0} \right) + \log^2\left( \frac{\tilde \omega_-}{\omega_0} \right)}.
}
This simple expression invites to investigate the optimal circuit in terms of the normal mode coordinates $\tilde x_\pm$. Such an observation can also be anticipated from the ground state wave function which is a factorized Gaussian state in the normal mode basis, i.e.
\eqn{
\psi_0(\tilde x_+,\tilde x_-) = \frac{(\twp \twm)^{1/4}}{\sqrt{\pi}} e^{ -\frac{1}{2} (\twp \tilde x_+^2 + \twm \tilde x_-^2) }.
}
To work out  the explicit form of the optimal circuit $U_0$ in normal mode basis, we first need to identify the required transformation matrix which yields the change $[\tilde x_+,\tilde x_-]^T = R_2 [x_1,x_2]^T $. It turns out that such a transformation can be realized via the following orthogonal rotation matrix 
\eqn{
R_2 = \frac{1}{\sqrt{2}} \begin{bmatrix}
1 & 1 \\ 
1 & -1
\end{bmatrix}\quad \text{with}\quad R R^T = R^T R = \mathbf{1},
}
where
\eqn{
\tilde A_\mathrm{T} = \begin{bmatrix}
\twp & 0\\ 
0 & \twm
\end{bmatrix},\qquad 
\tilde A_\mathrm{R} = \omega_0 \mathbf{1}
}
follow due to $\tilde A = R_2 A R_2^T$. Then, once the matrix $R_2$ is known, the optimal straight line circuit $U_0$ can be transformed to operate in the normal mode space via the transformation $\tilde U_0 (s) = R_2 U_0(s) R_2^T$. In contrast to $U_0$, the latter transformation results in a considerable simplification of the normal mode
circuit\footnote{Of course, the simple circuit $\tilde U_0$ can be obtained if the correct basis of generators in normal mode subspace are employed. Introducing the corresponding set of indices in that subspace, i.e. $\tilde I \in \{ ++,+-,-+,-- \}$, the generators $\tilde M_{\tilde I}$ are formally equal to $M_{\tilde I}$, but act in a different space. Using the relation 
\eqns{
\tilde M_{\tilde I} = R_2 M_{\tilde I} R_2^T \Rightarrow M_{\tilde I} = R_2^T \tilde{M}_{\tilde I} R_2
} 
the $\tilde M_{\tilde I}$ generators can be transformed to act on states expressed in position basis.  
Then, transforming \eqref{eqn:tU0} according to $R_2^T \tilde U_0(s) R_2$, where 
\eqns{
\tilde U_0(s) = \exp \left( \tilde Y^{\tilde I}(s) \tilde M_{\tilde I} \right),
}
leads to the optimal circuit from \eqref{eqn:U0}, i.e.
\eqns{
U_0(s) = \exp  \left( Y^{\tilde I}(s) M_{\tilde I} \right).
}
}
$\tilde U_0$ with the following diagonal form
\eqn{
\tilde U_0(s) = \exp\left(  \begin{bmatrix}
\frac{1}{2} \log\left(\frac{\twp}{\omega_0}\right) & 0 \\ 
0 &  \frac{1}{2} \log \left(\frac{\twm}{\omega_0}\right)
\end{bmatrix} s  \right).
\label{eqn:tU0}
}
It can be seen that $\tilde U_0$ does not possess any off-diagonal entries. In other words, in normal mode subspace, there is no entanglement introduced. This is in line with the mentioned factorized Gaussian shape for the states $\tilde \psi_\mathrm{R}$ and $\tilde \psi_\mathrm{T}$. The factorization turns out to be a substantial simplification when generalizing the results for the regulated field theory. Now, if the analogous computation is made by using the generators $\tilde M_{\tilde I}$ and the velocity vector $\tilde Y^{\tilde I}$ operating in the normal mode subspace, it follows that the complexity remains as in  \eqref{eqn:C2}, i.e. the complexity $\mathcal C_2$ for the $\mathcal F_2$ cost is independent of the basis. This feature can be comprehended if one uses the transformation matrix which transforms the generators $M_J$ to $M_{\tilde I}$.  Let us note that the former one, i.e. $M_J = R_2^T \tilde M_J R_2$, act in the physical basis to scale/entangle the coordinates $x_{1,2}$, whereas $\tilde M_{\tilde I} = R_2  M_{\tilde I} R_2^T$ scale/entangle the coordinates $x_{+,-}$ in the normal mode basis. The corresponding matrix is orthogonal and can be constructed according to the prescription \cite{Jefferson:2017sdb}
\eqn{
\widehat R_{\tilde I J} = R_{k a} \otimes R_{lb},\ \quad k,l \in \{ +,- \},\quad a,b \in \{ 1,2  \}
} 
such that $M_{\tilde I} = \widehat R_{\tilde I J} M_J$. Utilizing the matrix $\widehat R_{\tilde I J}$, the mentioned basis independence is nothing but the consequence of the equality
\eqn{
\left. \eqref{eqn:ds2} \right\vert_{G_{IJ} = \delta_{I J}}=\ &\delta_{\tilde I \tilde J}\mathrm{tr}(dU U^{-1} M_{\tilde I}^T)\  \mathrm{tr}(dU U^{-1} M_{\tilde J}^T).
}

Very similar to the discussion above, one can show that such a basis independence also holds for the $\mathcal F_{\kappa=2}$ cost set by
\eqn{
\mathcal F_{\kappa=2} \:= \sum_I G^I |Y^I|^2
}
where we will use $G^I = 1$ for the moment.
In this case, the complexity in both bases turns out to be
\eqnsplit{
\mathcal C_{\kappa=2} = 
\frac{1}{4} \left(  \log^2\left( \frac{\tilde \omega_+}{\omega_0} \right) + \log^2\left( \frac{\tilde \omega_-}{\omega_0} \right)   \right) 
\label{eqn:Ckappa=2}
}
i.e. $\mathcal C_{\kappa=2} = \mathcal C_{2}^2$. 
However, this basis independence does generally not occur for costs of the form
\eqn{
\mathcal F_{\kappa} = \sum_I G^I \vert Y^I \vert^\kappa\qquad \forall\ \kappa \neq 2.
} 
We should add that the cost $\mathcal F_{\kappa = 2}$ is of particular relevance for our studies, since it leads to the same leading UV divergence as found in holographic complexity computations. More details regarding this issue are discussed in section \ref{subsubsec:C-holo}.

\subsection{Lattice}
\label{subsec:Nosc}

\subsubsection{One dimensional}
The previous results can be generalized for the regulated field theory by a line lattice. The corresponding Hamiltonian is of the form
\eqn{
H = \frac{1}{2} \sum_{j=0}^{N-1} \left( p_j^2 + \omega^2 x_j^2 + \Omega^2(x_j  -  x_{j+1})^2 \right)
\label{eqn:H_Nosc}
}
which satisfies the periodic boundary condition $x_{j+N} = x_j$. It describes a discretized field placed on a circle with length \eqn{
L=N \delta. 
\label{eqn:L}
}
Motivated by the simplifications in normal mode basis, we rewrite $H$ in this representation, i.e.
\eqn{
H = \frac{1}{2} \sum_{k=0}^{N-1} \left( \vert \tilde p_k \vert^2 + \tilde \omega^2_k \vert \tilde x_k \vert^2 \right).
}
The normal mode coordinates can be deduced from a discrete Fourier transform,
\eqn{
\tilde x_k = \frac{1}{\sqrt{N}} \sum_{j=0}^{N-1} \exp \left(  \frac{2 \pi i k}{N} j \right) x_j,
}
where $k \in \{ 0,\ldots,N-1 \}$ and $\tilde x_k^\dag = \tilde x_{N-k}$. For $N=2$, we find 
the coordinates
\eqnsplit{
\tilde x_0 = \frac{1}{\sqrt{2}} (x_0 + x_1),\quad
\tilde x_1 = \frac{1}{\sqrt{2}} (x_0 - x_1)
} 
which of course correspond to the one introduced in \eqref{eqn:nm-variables} if we identify $\tilde x_{0,1} \leftrightarrow \tilde x_{+,-}$ and $x_{0,1} \leftrightarrow x_{1,2}$.
The normal mode frequencies are expressed in terms of the physical frequencies $\omega$ and $\Omega$,
\eqn{
\tilde \omega_k^2 = \omega^2 + 4 \Omega^2 \sin^2 \left(  \frac{\pi k}{N} \right).
}
Note that, in contrast to the system of two coupled harmonic oscillators, cf. equation \eqref{eqn:nm-variables}, here we have a factor $4$ in front of the second term.
This goes back to the periodic boundary condition.
This can be also seen when we set $N=2$ in \eqref{eqn:H_Nosc} and demand $x_2 \equiv x_0$ such that the term being proportional to $\Omega^2$ becomes doubled, see also section \ref{subsubsec:N=2-unpen}.

In normal mode basis, the factorized Gaussian shape of both the reference state and target ground state is preserved. Therefore, the previous studies for $N=2$ can be generalized to the case of $N$ oscillators. In total, one gets $N^2$ generators which give rise to $N \times N$ matrices. This extends the initial group $GL(2,\mathbb{R})$ to $GL(N,\mathbb{R})$. For the $\mathcal F_2$ cost from \eqref{eqn:F2}, an appropriate parameterization spanned with $N^2$ coordinates of a general element $U \in GL(N, \mathbb{R})$ is needed. Accordingly, the optimal circuit will then be determined by the optimal geodesic in this extended geometry characterized by the corresponding right-invariant metric $ds^2$. 

Following the usual procedure can be very challenging due to the large number of coordinates parameterizing the extended metric. Instead, one may expect that, similar to the previously discussed simplified diagonal form of the optimal circuit in normal mode basis, a simplification of such type also applies for the present problem. Hence, the most optimal circuit  would simply amplify the Gaussian width for each of the normal mode coordinates. In particular, the extended optimal circuit would not introduce any entanglement between the normal modes. Indeed, it has been shown that the optimal circuit takes the form of the generalized straight line circuit
\eqn{
\tilde U_0(s) = \exp \left( \tilde M_0 s \right)
\label{eqn:tilde U_0(s)-N}
}
where 
\eqn{
\tilde M_0 = \text{diag} \left[\frac{1}{2} \log \left( \frac{\tilde \omega_0}{\omega_0} \right), \ldots,  \frac{1}{2} \log \left( \frac{\tilde \omega_{N-1}}{\omega_0} \right) \right].
}
For the details showing that $\tilde U_0$ from \eqref{eqn:tilde U_0(s)-N} indeed corresponds to the optimal circuit, we refer the interested reader to \cite{Jefferson:2017sdb}.
Bringing all together, the complexity for the one dimensional regulated field theory becomes
\eqn{
\mathcal C_2 = \frac{1}{2} \sqrt{  \sum_{k=0}^{N-1}     \log^2 \left( \frac{\tilde \omega_k}{\omega_0} \right) }.
\label{eqn:C2-N-gen}
}

\subsubsection{$d$ dimensional}
\label{subsubsec:d-dim}
The complexity for the one dimensional lattice can be easily extended to the $(d-1)$ dimensional lattice consisting of $N^{d-1}$ oscillators. The final expression is of the from
\eqn{
\mathcal C_2 = \frac{1}{2} \sqrt{  \sum_{\{k_j\}=0}^{N-1}     \log^2 \left( \frac{\tilde \omega_{\vec{k}}}{\omega_0} \right) },
\label{eqn:C2-N-gen-ddim}
}
where $k_j$ denote the momentum vector components. The corresponding expression for the normal mode frequencies reads
\eqn{
\tilde \omega_{\vec{k}}^2 = \omega^2 + 4 \Omega^2 \sum_{j=1}^{d-1} \sin^2 \left(  \frac{\pi k_j}{N} \right).
\label{eqn:nm-freqs-ddim}
}

\subsection{Comparison with holographic complexity}
For the comparison with holographic complexity, it is advantageous to express the complexity from \eqref{eqn:C2-N-gen-ddim} in terms of the field theory parameters. Then, the normal mode frequencies introduced in \eqref{eqn:nm-freqs-ddim} become
\eqn{
\tilde \omega_{\vec{k}}^2 = m^2 + \lc \frac{2}{\delta} \rc^2  \sum_{j=1}^{d-1}   \sin^2 \left(  \frac{\pi k_j}{N} \right)
\label{eqn:nm-freqs-ddim-rep}
}
according to the definitions in \eqref{eqn:qft-parameters}.
An estimation for the complexity \eqref{eqn:C2-N-gen-ddim} in terms of the volume of the system, $V$, is also very useful. This can be accomplished if we first write 
$V = L^{d-1}$ by assuming an equisided lattice. Afterwards, applying the relation \eqref{eqn:L}, the total number of oscillators can be expressed as
\eqn{
N^{d-1} = \frac{V}{\delta^{d-1}}.
\label{eqn:N}
}
By using the relation in \eqref{eqn:N}, it can be shown that the complexity scales as
\eqn{
\mathcal C_2 \sim \frac{N^{\frac{d-1}{2}}}{2} \log \left( \frac{\tilde \omega_{\vec{k}}}{\omega_0}  \right).
\label{eqn:C2-approx}
}

\subsubsection{UV divergence in QFT}
\label{subsubsec:UV-div-QFT}
The leading order contribution to complexity in QFT is determined by the UV modes. We can take $\delta$ to be the UV cutoff where $\delta m \ll 1$. In this UV approximation, we may estimate the normal mode frequencies in \eqref{eqn:nm-freqs-ddim-rep} by assuming
\eqn{
\tilde \omega_{\vec{k}} \sim 1/\delta.
\label{eqn:uv-approx}
}
Using the approximation above, as well as the expressions in \eqref{eqn:N} and \eqref{eqn:C2-approx}, the leading order contribution to the complexity depending on the UV cutoff scales as
\eqn{
\mathcal C_2 
\sim \sqrt{ \frac{V}{\delta^{d-1}}  }.
}
Remember that for the two coupled oscillators we have seen that $\mathcal C_{\kappa=2} = \mathcal{C}_2^2$.
Taking this relation into account, we may immediately get the leading order UV divergence on the lattice for the $\mathcal F_{\kappa = 2}$ cost which scales as
\eqn{
\mathcal C_{\kappa=2} \sim \frac{V}{\delta^{d-1}}.
\label{eqn:uv-kappa2-approx}
}
\subsubsection{UV divergence in holography}
\label{subsubsec:C-holo}
Studies of the UV divergence in holographic complexity based on the CV and CA proposals 
have shown that the leading contribution scales as  \cite{Carmi:2016wjl}
\eqn{
\mathcal C_\text{V,A} \sim \frac{V}{\delta^{d-1}}.
\label{eqn:C-V-A}
}
This behavior is similar to the scaling for the $\mathcal F_{\kappa =2 }$ cost from \eqref{eqn:uv-kappa2-approx}. Apart from that, it has been argued that the following depth function for the circuit complexity
\eqn{
\tilde{\mathcal{D}}_\kappa = \int_0^1 ds\ \sum_{\tilde I} \vert Y^{\tilde I}(s) \vert ^\kappa
} 
with $\kappa \geq 1$ would give rise to the same UV divergence as in \eqref{eqn:C-V-A}. On the field theory side we simply get the mentioned scaling,
\eqn{
\mathcal C_\kappa \sim \frac{V}{\delta^{d-1}} \left| \logb{\frac{1}{\delta \omega_0}} \right|^\kappa,
\label{eqn:C_kappa}
} 
however, with an additional logarithmic factor which will be discussed in detail below.
The reason for the same prefactor for all $\kappa$ lies in the fact that for the depth function in normal mode basis, i.e. $\tilde{\mathcal{D}}_\kappa$, the previously discussed straight line circuit still corresponds to the optimal circuit \cite{Jefferson:2017sdb}. This basis independence for all $\kappa$ including the $\mathcal F_2$ cost is the reason why the optimal geodesic remains the same. In the original position basis, i.e. ${\mathcal{D}}_\kappa$, this is generally not the case. The only exceptions are the $\mathcal F_{\kappa=2}$ and $\mathcal F_2$ cost functions.

For further interesting similarities in the divergence structure, we can compare with the leading order contribution in the CA proposal,
\eqn{
\mathcal C_\text{A} \sim \frac{V}{\delta^{d-1}} \logb{\frac{L_\text{AdS}}{\Lambda \delta}}
\label{eqn:C-CA}
}
where $L_\text{AdS}$ is the curvature scale of the AdS bulk spacetime, $\Lambda$ is an arbitrary dimensionless coefficient fixing the null normals on the WDW patch boundary and $\delta$ is the short distance cutoff scale in the boundary CFT \cite{Carmi:2016wjl}. As suggested in \cite{Jefferson:2017sdb}, for eliminating $L_\text{AdS}$ from the expression---since $\mathcal C_\text{A}$ is expected to be defined in the boundary CFT which therefore should not depend on any bulk AdS scale---one may fix the dimensionless coefficient as $\Lambda = \omega_0 L_\text{AdS}$. Of course, $\omega_0$ is still an arbitrary frequency. However, with the latter replacement, the complexity simply reduces to
\eqn{
\mathcal C_\text{A} \sim \frac{V}{\delta^{d-1}} \logb{\frac{1}{\delta \omega_0}}.
\label{eqn:C-CA-replaced}
}
It is remarkable that this expression is similar to the QFT expression in \eqref{eqn:C_kappa} which also depends on some unspecified frequency $\omega_0$ of the reference state. The found coincidence may therefore suggest an interesting relation between the notions of complexity defined in holography and QFT.

\section{Penalized geometry I: Coupled oscillators}
\label{sec:pen-osc}
\begin{figure}[h!]
  \centering
    \includegraphics[width=0.48\textwidth]{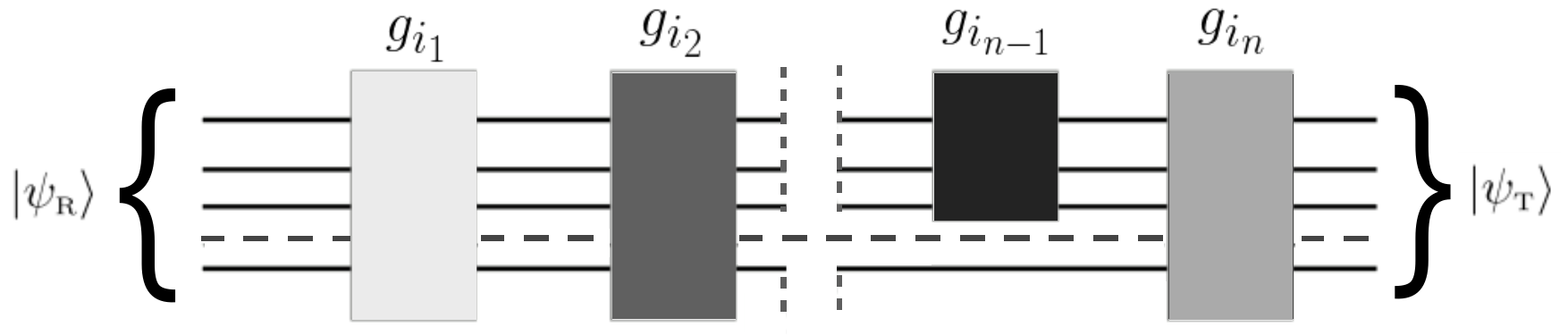}
      \caption{A quantum circuit constructed by weighted gates. The color scaling for the gates shall indicate the difference in the gate weighting.}
      \label{fig:circuit-weighted}
\end{figure}

So far, both the scaling and entangling gates have been treated equally, i.e. each of them received the same cost when constructing the optimal circuit. 
As pointed out in the introduction, technically, for implementing the notion of locality one would need to introduce some weighting factors which vary the strength of the acting gates according to the separation between the corresponding points. Such a situation is sketched in figure \ref{fig:circuit-weighted}.

However, recalling our motivation in the introduction, we may consider a simpler setup via penalizing the gates uniformly. 
In the following, we first briefly discuss these effects
for the system of two coupled harmonic oscillators subjected to the $\mathcal F_2$ cost as already explored in \cite{Jefferson:2017sdb}. Afterwards, we extend our discussion for $\mathcal F_{\kappa=2}$ and work out the differences. The results constitute the basis for the field theory computations later in section \ref{sec:pen-qft}.

\subsection{$\mathcal{F}_2$ cost}
Introducing weighting factors for the entangling gates simply means modifying the off-diagonal directions in the circuit. To do so, we may write the tensor $G_{I J}$ in \eqref{eqn:circ-depth} as
\eqn{
G_{I J} = \text{diag} [1,\alpha^2,\alpha^2,1]
\label{eqn:GI-pen}
}
with $\alpha^2 > 1$. Of course, for $\alpha = 1$ one would arrive at the original situation in which no extra cost is assigned to the entangling gates. With the extended penalty tensor from \eqref{eqn:GI-pen}, the corresponding metric would be modified due to the appearance of additional mixing terms. The final expression may principally be treated as usual. Namely, after identifying  the corresponding Killing vectors, one would need to find all conserved charges in order to solve the geodesic equations. 
However, writing down the full metric, it turns out that this task is too complicated to be performed. This problem has been avoided by assuming $\alpha \gg y_1,\rho_1 \gg 1$ and neglecting some components of the metric to simplify the full expression. 
In this way, it is possible to construct a kind of segmented path which is not a geodesic, but is capable to come very close to the optimal geodesic \cite{Jefferson:2017sdb}.
The segmented circuit constructed by exploiting the described approximate path takes the form
\eqn{
U_p(s) \simeq \left\{ 
\begin{array}{cc}
U_{p_a}(s) ,& 0 \leq s \leq \frac{1}{2} \\\\
U_{p_b}(s) ,&  \frac{1}{2} \leq s \leq 1
\end{array}.
\right.
}
The circuits for the two segments are determined by
\eqnsplit{
U_{p_a}(s) &=  \begin{bmatrix}
e^{-\rho_1 2 s} & 0 \\ 
0 & e^{\rho_1 2 s}
\end{bmatrix}
}
and
\eqn{
U_{p_b}(s) &= e^{y_1(2s-1)}  \begin{bmatrix}
\cos( \phi) e^{-\rho_1} & -\sin( \phi) e^{\rho_1} \\ 
\sin( \phi) e^{-\rho_1} & \cos( \phi) e^{\rho_1}
\end{bmatrix}
}
where we have defined $\phi := \frac{\pi}{4}(2s-1)$. Instead of continuing with these expressions, we want to rewrite them in exponential form. This can be easily done by utilizing the relation \eqref{eqn:Y_I-Mat} which yields the desired exponential forms
\eqn{
U_{p_a}(s) = \exp \left( \begin{bmatrix}
-\rho_1 & 0 \\ 
0 & \rho_1
\end{bmatrix}
2s
\right)
\label{eqn:U_pa}
}
and
\eqn{
U_{p_b}(s) = \exp \left( \begin{bmatrix}
y_1 & \frac{\pi}{4} \\ 
\frac{\pi}{4} & y_1
\end{bmatrix}
\frac{s}{2}
\right).
\label{eqn:U_pb}
}
The expressions above turn out to be much simpler and more suitable for later use.
Now, using \eqref{eqn:U_pa} and \eqref{eqn:U_pb} in combination with  \eqref{eqn:path-ordered-Mat}, we can easily find the velocity vector components.  Note that the generators are the one which are given in \eqref{eqn:M_ij}. Once the components of $Y^I(\tilde s)$ are found, we can use the splitting
\eqn{
\mathcal C = \mathcal D(U_p) \simeq \mathcal D(U_{p_a}) + \mathcal D(U_{p_b})
\label{eqn:C-split}
}
to deduce the complexity. The circuit depths on the right-hand side of \eqref{eqn:C-split} can be computed according  to the integral form introduced in \eqref{eqn:circ-depth}. Note that the latter is associated with the $\mathcal F_2$ cost.

We should add that the given segmented construction can also be used for the $\mathcal F_{\kappa=2}$ cost. Differently, for a general $\kappa \geq 1$, it does not yield a circuit depth which comes close to the optimal one. One might argue that changing the basis and computing $\tilde{\mathcal D}_\kappa$ instead of $\mathcal D_\kappa$ may solve this problem. However, it has been discussed that, once the geometry is penalized, the depth function will not be invariant under a basis change  \cite{Jefferson:2017sdb}. Let us continue with the complexity for the $\mathcal F_2$ cost. The final result, which can be obtained by plugging the mentioned segmented path into the normalization constant,\footnote{The normalization constant gives the length of the geodesic. It corresponds to the complexity if the geodesic is the minimal one. 

For $\mathcal F_2$, the normalization constant is simply given by the integral 
\eqns{
\mathcal D(U) = \int_0^1 ds\ \sqrt{ g_{ij} \dot x^i(s) \dot x^j(s) }
}
where $\pmb{x}(s) = [\tau(s),\rho(s),\theta(s),y(s)]$. 

Here, $s$ is the scaled affine  parameter. The components of $\pmb{x}$ are the mentioned parameterization variables in section \ref{subsubsec:pos-space} for the subgroup $SL(2,\mathbb{R})$ and the $\mathbb{R}$ fibre.
}
takes the form
\eqnsplit{
\pD_{2}(U_p) &\simeq \sqrt{2} \rho_1  + \sqrt{2y_1^2 +2 \alpha^2 \left( \frac{\pi}{4} \right)^2 }\\
&\simeq \sqrt{2} \rho_1 +  \frac{\pi}{2 \sqrt{2}} \alpha + \mathcal{O}\left( \frac{1}{\alpha^2} \right).
}
Here, the superscript $p$ indicates that we work in penalized geometry. Now, if we go back and compare this result with the unpenalized straight line circuit depth in \eqref{eqn:C2}, we find that
\eqn{
\pD_{2}(U_p) \gg \D_{2}(U_0)
}
which follows due to the mentioned assumption $\alpha \gg \rho_1,y_1 \gg 1$. 
Being in agreement with the naive expectation, the complexity increases when a higher cost is assigned to the entangling gates.

\subsection{$\mathcal F_{\kappa = 2}$ cost}
Once the exponential forms in \eqref{eqn:U_pa} and \eqref{eqn:U_pb} are known, finding the complexity in penalized geometry for the $\mathcal F_{\kappa = 2}$ cost becomes straightforward. Similar to the previous case, we first fix  
\eqn{
G^I = [1,\alpha^2,\alpha^2,1]^T.
}
Then, by using the depth integral
\eqn{
\mathcal D_{\kappa=2} = \int_0^1 d\tilde s\ \sum_I G_{I} |Y^I(\tilde s) |^2
}
and the splitting in \eqref{eqn:C-split}, we obtain
\eqn{
\pD_{\kappa=2}(U_p) \simeq 2 \rho_1^2 + 2y_1^2 + 2\alpha^2 \left( \frac{\pi}{4} \right)^2.
\label{eqn:C_kappa2_approx}
}
As before, we can compare the result \eqref{eqn:C_kappa2_approx} with the complexity in unpenalized geometry, i.e. $\D_{\kappa = 2}(U_0) = \D_{2}^2(U_0)$.
The difference between both complexities becomes
\eqn{
\pD_{\kappa=2}(U_p)  - \D_{ \kappa=2}(U_0) \simeq 2\alpha^2 \left( \frac{\pi}{4} \right)^2 \gg 1
}
showing again that assigning a higher cost to the entangling gates increases the complexity.

\section{Penalized geometry II: Field theory}
\label{sec:pen-qft}
In this section, we study the effect of introducing weighting factors for the entangling gates in the regulated field theory placed on the lattice. As we have stressed above, for the two coupled oscillators, a direct attempt via analytically finding the minimal geodesic in full penalized geometry is already very challenging.
However, certain assumptions allowed to simplify the problem under consideration. Even though the exact geodesic could not be analytically derived for the simplified case as well, an approximate segmented path has been constructed which comes very close to the optimal geodesic. This has been explicitly illustrated by comparing the corresponding complexities \cite{Jefferson:2017sdb}.

For the generalization on the lattice, the question is, how to find the the optimal geodesic in the extended geometry where a higher cost is attributed to the entangling gates? For instance, one may argue that by changing to the normal mode basis one could consider a perturbation around the previous straight line path, since this particular solution yields the optimal geodesic in unpenalized geometry for all $\mathcal F_\kappa$ costs where $\kappa \geq 1$. In general, this strategy will not simplify the problem either, since introducing a penalty tensor in one basis does not automatically lead to the same penalty tensor when working in a different basis. However, for certain choices such as the $\mathcal F_2$ and $\mathcal F_{\kappa=2}$ costs, as we will see, the straight line circuit may indeed be used in normal mode basis. Let us bring to mind that for these the complexity is basis independent if no extra cost is assigned to the entangling gates, see section \ref{sec:setup}.

In the following, we use the results already discussed in the unpenalized case for deriving the complexity without solving the geodesic equations explicitly. For clarifying the underlying procedure, we start again with the simple case of two coupled oscillators. Please note, that all the following steps are described for the $\mathcal F_2$ cost. But we know from earlier discussions that the findings can also be applied to the $\mathcal F_{\kappa = 2}$ cost which will be our main choice later.

First, we begin by assuming that $U_0$ is the optimal circuit which implements the following operation
\eqn{
\vert \psiT \rangle = U_0 \vert \psiR \rangle
}
in unpenalized geometry.
The corresponding complexity shall be denoted by $\mathcal{C}(\psiT , \psiR \vert U_0)$. In addition, we assume that the analogous optimal circuit $U_*$ in penalized geometry (i.e. when the entangling gates are weighted as above) is known as well, which implements
\eqn{
\vert \psiT \rangle = U_* \vert \psiR \rangle.
}
For the latter, the complexity shall be denoted by $\mathcal{C}(\psiT , \psiR \vert U_*)$. 
As previously discussed, we know that both complexities now satisfy the following inequality
\eqn{
\mathcal{C}(\psiT , \psiR \vert U_*)
>
\mathcal{C}(\psiT , \psiR \vert U_0).
}
Next, let us assume that we construct an artificial target state $\psiT^*$ which satisfies the implementation
\eqn{
\vert \psiT^* \rangle = U_0 \vert \psiR \rangle
\label{eqn:des-eqn}
}
in unpenalized geometry, but yields a different complexity $\mathcal{C}(\psiT , \psiR \vert U_*)$. 
Notice that the target state $\vert \psiT \rangle$ depends on the physical parameters $\omega$ and $\delta$ by construction. 
Now, if we keep the frequency $\omega$ unchanged and demand \eqref{eqn:des-eqn}, it is possible to fulfill such an implementation if we introduce a modified parameter $\delta_*$ which is assigned to the mentioned state $\vert \psiT^* \rangle$. 

Then, we may replace the original cutoff scale $\delta$ in the optimal circuit $U_0$ by $\delta_*$. This will simply allow deriving the complexity in the penalized geometry by using the optimal geodesic in the unpenalized geometry. 
The procedure works for the states under consideration, since both are taken to be positive quadratic forms.
As we will show soon, such a replacement will regulate the desired change in complexity. 

It is important to note that this replacement is rather technical. The cutoff scale in the theory will still be determined by $\delta$.

\subsection{Modified parameter}
\label{subsec:eff-cutoff}
In the following, we explicitly illustrate the described strategy from above for the system of two coupled oscillators subjected to the $\mathcal F_{\kappa = 2}$ cost. 
For reasons which will become clear below, we cannot use the expression in \eqref{eqn:C_kappa2_approx} for the optimal circuit depth in the penalized geometry which we again denote by $\pD_{\kappa = 2}$. Recall that the result in \eqref{eqn:C_kappa2_approx} is based on the segmented path approach which is only valid when the weighting factor introduced in \eqref{eqn:GI-pen} satisfies $\alpha \gg y_1,\rho_1 \gg 1$. A more general expression where the range of $\alpha$ is not restricted from below has been worked out for the $\mathcal F_2$ cost  \cite{Jefferson:2017sdb}. According to our previous findings, we may also use this expression in the present case by simply squaring it. Proceeding in this way yields the desired circuit depth which takes the form
\eqnsplit{
\pD_{\kappa = 2} &= 2 \left( \frac{\alpha}{2} \tan^{-1} \left(  \sqrt{\alpha^2 - 1}\right) + \rho_1  \right)^2 + 2y_1^2.
\label{eqn:Dp_kappa2}
}
Of course, for $\alpha \rightarrow 1$ it reduces to the straight line circuit depth $\D(U_0)$ which equals to the square of \eqref{eqn:C2}.

Having specified this, what remains to be done is finding the solution $\delta_*$ which solves the equation $\D_{\kappa = 2} = \pD_{\kappa = 2}$.
Note that the right-hand side of the latter equation has to be written in terms of the functions $y_1(\delta,\omega,\omega_0)$ and $\rho_1(\delta,\omega) $
where we recall from \eqref{eqn:y1-rho1-nm} that
\eqnsplit{
y_1(\delta,\omega,\omega_0) &= \frac{1}{4} \logb{ \frac{\omega^2}{\omega_0^2} \sqrt{1 + \frac{2}{\delta^2 \omega^2} }},\\
\rho_1(\delta,\omega) &= \frac{1}{4} \logb{ \sqrt{1 + \frac{2}{\delta^2 \omega^2} }}.
\label{eqn:y1rho1-exp}
}
Contrary to that, the functions $y_1$ and $\rho_1$ on the left-hand side have to be written in terms of $\delta_*$. Finally, using the square of the straight line circuit depth in \eqref{eqn:C2}, the explicit equation we need to solve takes the form
\eqnsplit{
2 y_1^2(\delta_*,\omega,\omega_0)  + 2 \rho_1^2(\delta_*,\omega) =  2 y_1^2(\delta,\omega,\omega_0)
+ 2 \left( \frac{\alpha}{2} \tan^{-1} \left(  \sqrt{\alpha^2 - 1}\right) + \rho_1(\delta,\omega)  \right)^2.
}
This equation can be analytically solved. An appropriate treatment yields the following result
\eqn{
\delta_* =  \sqrt{2} \left(  \omega_0^2  e^{  \sqrt{  \chi(\alpha,\omega,\delta) + \log^2\left(  \frac{2 + \delta^2 \omega^2}{\delta^2 \omega_0^2 }  \right)  } }   - \omega^2  \right)^{-1/2}
\label{eqn:dstar-ana}
}
where we have defined
\eqnsplit{
\chi(\alpha,\omega,\delta) := 4^2 \lc \alpha \tan^{-1} \left(  \sqrt{\alpha^2-1} \right) \rc
\left( \frac{\alpha}{2}  \tan^{-1} \left(  \sqrt{\alpha^2-1} \right) + 2 \rho_1(\omega,\delta) \right)
\label{eqn:chi-ana}
}
due to  practical reasons.  
The expression in \eqref{eqn:dstar-ana} is our key result which will be important in the remainder of this section.
Of course, once we take the unpenalized limit, the only $\alpha$ dependent part in \eqref{eqn:dstar-ana} vanishes, i.e. $\chi \rightarrow 0$, and we end up with the original cutoff scale $\delta$,
\eqn{
\lim_{\alpha \rightarrow 1} \delta_* = \delta.
\label{eqn:lim-delta*}
}

\subsection{Small penalty} 
\label{subsec:small-pen}
To highlight the effect of assigning a higher cost to the entangling gates, it is advantageous to use the perturbative expansion of $\delta_*$ for small penalties. This can be achieved for moderate $\alpha$. More precisely,
the prefactor in front of $2 \rho_1$ in \eqref{eqn:chi-ana} which only depends on the weighting factor has to be sufficiently small, i.e. $\chi \ll 1$. This can be controlled with a weighting factor that is $\alpha \neq 1$, but close enough to one.
In this case, we may perturb $\delta_*$ around $\chi = 0$ which yields 
\eqnsplit{
\delta_* 
&\simeq \delta \lc  1 -   R  \chi  +  \mathcal{O}(\chi^{ 2 }) \rc 
\label{eqn:delta-taylor}
}
where we have defined 
\eqn{
R \equiv \frac{e^{\sqrt{K}}  (\omega_0 \delta)^2  }{2^3 \sqrt{K}},\quad 
K \equiv \log^2 \lc \frac{2 + \delta^2 \omega^2}{\delta^2  \omega_0^2 }  \rc.
\label{eqn:R-and-K}
}
Suppose that we fix the reference frequency $\omega_0 \sim e^{-\sigma} / \delta $ at some scale $\sim 1/\delta$. The overall numerical parameter $\sigma$ ensures that $\omega_0 > 1/\delta$. Furthermore, we would like to work in the UV approximation described in section \ref{subsubsec:UV-div-QFT}. Using the corresponding assumptions, we may estimate 
$K \sim - \log^2 \lc \delta \omega_0 \rc$.
Observe that the dimensionless factor $R$ in \eqref{eqn:R-and-K} is positive in this regime. Also, $\chi$ is taken to be sufficiently small as described above. Therefore, we find that the expression in the brackets from \eqref{eqn:delta-taylor} decreases as soon as the weighting factor $\alpha$ becomes different from one, i.e. $\delta_* < \delta$. This change increases the complexity due to 
\eqnsplit{
y_1(\delta_*,\ldots) &> y_1(\delta,\ldots),\\
\rho_1(\delta_*,\ldots) &> \rho_1(\delta,\ldots),
} 
see \eqref{eqn:y1rho1-exp}. This trend is for sure expected according to the nonperturbative expression from \eqref{eqn:Dp_kappa2}. 

Up to a numerical factor $\mathcal N$ multiplied with $R$ from \eqref{eqn:R-and-K}, which is not relevant for our discussion later (thus we set $\mathcal N \equiv 1$), we may also write an expansion of the following form 
\eqn{
\delta_* \simeq \delta \lc  1 -  \underline R  \tilde{\alpha} +  \mathcal{O}(\tilde{\alpha}^2)  \rc, \quad \underline{R} :=  \sqrt{K} R
\label{eqn:delta-taylor-alpha}
}
where $\tilde{\alpha} := \sqrt{\alpha - 1}$. Writing $\delta_*  \simeq \delta (\ldots)$ as in \eqref{eqn:delta-taylor-alpha} will allow to make observations more comprehensible in certain cases.
\subsection{On the lattice}

\subsubsection{$N=2$, unpenalized}
\label{subsubsec:N=2-unpen}
For the moment, let us consider a periodic lattice with $N=2$. The related Hamiltonian describing this system takes the following form
\eqn{
H = \frac{1}{2} \lc p_0^2 + p_1^2 + \omega^2 (x_0^2 + x_1^2) + \frac{2}{\delta^2} (x_0 - x_1)^2 \rc
\label{eqn:H-2osc-periodic}
}
where we have imposed the boundary condition $x_2 \equiv x_0$. 
Next, we write the Hamiltonian from \eqref{eqn:H-2osc-periodic} in terms of 
\eqn{
\bar \delta := \delta / \sqrt{2}
\label{eqn:replace-check}
} 
which just gives rise to the same Hamiltonian as introduced in \eqref{eqn:H-2osc}. 
Recall that according to our previous discussion in section \ref{subsec:eff-cutoff}, the optimal circuit depth $\D_{\kappa=2}$ from \eqref{eqn:Ckappa=2} satisfies the relation
\eqnsplit{
2 y_1^2(\bar \delta,\omega,\omega_0) + 2  \rho_1^2(\bar \delta,\omega)
= \frac{1}{4} \sum_{k=0}^1 \log^2 \lc \frac{\tilde \omega_k}{\omega_0}  \rc
\label{eqn:sl-N2}
}
where the corresponding normal mode frequencies are given as
\eqn{
\tilde \omega_0^2 = \omega^2,\quad \tilde \omega_1^2 = \omega^2 + 2/\bar \delta^2.
}
Now, based on the discussion in section \ref{subsubsec:d-dim}, it is straightforward to generalize these results for the $d$ dimensional case.
For doing so, we may write 
\eqn{
V = (N \delta)^{d-1},
}
although we should note that $N$ has been fixed above. Then, working in the UV approximation, i.e. $\tilde \omega_{\vec{k}} \rightarrow 1/{\bar \delta}$, we end up with the following leading order contribution to the complexity
\eqn{
\D_{\kappa=2} \sim \frac{V}{\delta^{d-1}}  \log^2 \lc  \frac{1}{\bar \delta \omega_0} \rc.
\label{eqn:C-N2}
} 
Of course, the expression in \eqref{eqn:C-N2} has been derived in unpenalized geometry.

\subsubsection{$N=2$, penalized}
\label{subsubsec:N=2-pen}
The previous findings can be easily extended to the penalized case when $\alpha > 1$. We have seen that the periodic lattice with $N=2$ reduces to the system of two coupled oscillators if $\bar \delta$ from \eqref{eqn:replace-check} replaces the original cutoff scale $\delta$. Thus, we may use the previous straight line solution and insert $\sqrt{2} \bdelta_* = \delta_*(\delta \rightarrow \sqrt{2} \delta)$ into the corresponding places. In the mentioned UV approximation,
this would simply mean that 
$\tilde \omega_{\vec{k}} \rightarrow 1/{\bdelta_*}$. 
Note that according to \eqref{eqn:lim-delta*} we get 
$\lim_{\alpha \rightarrow 1} \bdelta_* = \delta$.
The complexity in the penalized geometry then takes the form
\eqn{
\pD_{\kappa=2} \sim \frac{V}{\delta^{d-1}}  \log^2 \lc  \frac{1}{\bdelta_* \omega_0} \rc.
\label{eqn:C-N2-pen-nonper}
}
As in the preceding discussion, this expression has been derived by fixing $N = 2$. Even though \eqref{eqn:C-N2-pen-nonper} provides useful information on how (uniformly) weighting a certain class of gates (here, these are the entangling gates) would influence the complexity in the regulated field theory, we still need to extend the results for an arbitrary number of oscillators $N$. In the subsequent part, it will be argued that the relation from \eqref{eqn:C-N2-pen-nonper} also applies in the general case for any $N$.

As in the unpenalized case, let us close this part with expanding the $\alpha$ dependent parameter $\bdelta_*$ for small penalties which yields 
\eqn{
\bdelta_* \simeq \delta \lc  1 - \breve R \chi  + \mathcal{O}(\chi^2) \rc
}
where we have defined
\eqnsplit{
\breve R := \frac{e^{\sqrt{\breve{K}}}  (\omega_0 \delta)^2  }{2^{2} \sqrt{\breve{K}}},\quad \breve{K} := \log^2 \lc \frac{1 + \delta^2 \omega^2}{\delta^2  \omega_0^2 }  \rc.
}
Analogous to the expression in \eqref{eqn:delta-taylor-alpha}, we may also write an approximation of the form
\eqn{
\bdelta_* \simeq \delta \lc  1 - \underline{\breve R} \tilde{\alpha} + \mathcal{O}(\tilde{\alpha}^2) \rc, \quad \underline{\breve R} := \sqrt{\breve K} \breve R.
}
The leading order contribution to the complexity in the small $\tilde{\alpha}$ limit then becomes
\eqn{
\pD_{\kappa=2} \sim \frac{V}{\delta^{d-1}}  \log^2 \lc  \frac{1}{\delta \omega_0(1 - \underline{\breve R} \tilde{\alpha})} \rc.
\label{eqn:C-N2-pen-per}
}

\subsubsection{Arbitrary $N$, penalized}
We begin by noticing that the complexity in the unpenalized case is basis independent for the $\mathcal F_2$ cost, i.e.
\eqn{
ds^2 = \delta_{I J} dY^I \lc dY^J \rc^* = \delta_{\tilde I \tilde J} dY^{\tilde I} \lc dY^{\tilde J} \rc^* .
}
In general, introducing a penalty tensor, i.e. $\delta_{I J} \rightarrow G_{I J}$, makes the metric depending on the choice of the basis. That means, choosing a specific metric $G_{I J}$ in the original position basis would result in a different metric $G_{\tilde I \tilde J}$ in the normal mode basis. For the present setup, where the entangling gates are uniformly penalized with the weighting factor $\alpha^2$, it can be shown that 
\eqn{
G_{\tilde I \tilde J} = \widehat R_{\tilde I J} G_{I J} \widehat R_{I \tilde J}^T
}
yields a matrix which takes the form
\eqn{
(1+\alpha^2) \mathbf{1} + G_{\text{off},\tilde I \tilde J},
\label{eqn:1+Goff}
}
see e.g. \cite{Jefferson:2017sdb}.
The matrix $G_{\text{off},\tilde I \tilde J}$ is symmetric and only consists of off-diagonal entries of the form $1-\alpha^2$ which do not distinguish between the different classes of gates. We should add that $\widehat R_{\tilde I J}$ denotes the generalized matrix with respect to the generators of $GL(N,\mathbb{R})$, see section \ref{subsubsec:nm-space}.
So, the expression in \eqref{eqn:1+Goff} shows that in normal mode basis there is no extra cost attributed to the entangling gates relative to the scaling gates.
Moreover, the dominant contribution to the metric $G_{\tilde I \tilde J}$ is 
determined by $\delta_{I J}$ multiplied with the prefactor $1+\alpha^2$. 
Note that, except the difference due to the appearance of the prefactor, this is similar as in the unpenalized geometry where $G_{I J} = \delta_{I J}$. 

In the unpenalized case both the ground state $\psiT$ and the reference state $\psiR$ turn out to be factorized Gaussian states in normal mode basis, i.e.
\eqnsplit{
\psiR(\tilde x_k) &= \lc \frac{\omega_0}{\pi} \rc^{N/4} \expb{  -\frac{1}{2} \tilde x^\dag \tilde A_\text{R} \tilde x},\\
\psiT(\tilde x_k) &= \prod_{k=0}^{N-1} \lc  \frac{\tilde \omega_k}{\pi} \rc^{1/4} \expb{ -\frac{1}{2} \tilde x^\dag \tilde A_\text{T} \tilde x  },
}
with $ \tilde A_\text{R} = \omega_0 \mathbf{1}$, $\tilde A_\text{T} = \mathrm{diag}[\tilde \omega_0,\ldots,\tilde \omega_{N-1}]$ and  $\tilde x = R_N x$ where $R_N$ denotes the unitary matrix generalizing $R_2$ introduced in section \ref{subsubsec:nm-space}.
Such a factorization simplifies the optimal circuit $\tilde U_0 = R_N U_0 R_N^\dag$ in form of a diagonal matrix which amplifies each of the diagonal entries in the quadratic forms. As a consequence, the resulting geometry of the normal mode subspace remains flat, since the generators $M_{\tilde I} = \widehat R_{\tilde I J} M_J$ which construct the optimal circuit commute with one another. This property explains why the straight line solution yields the optimal geodesic for the regulated field on the lattice.

Similarly, we may also expect that the optimal circuit in the penalized geometry is a diagonal matrix. This goes back to observation that the metric in normal mode basis has the same shape as in the unpenalized case, i.e. $G_{\tilde I \tilde J} \sim \delta_{\tilde I \tilde J}$. Thus, we may again use the straight line solution. The additional change for $\alpha > 1$ can be incorporated by making the replacement $\delta \rightarrow \bdelta_*$ as illustrated in section \ref{subsubsec:N=2-pen}. 
Everything so far has been discussed for the $\mathcal F_2$ cost.
From earlier studies we know that the latter aspects also apply for the $\mathcal F_{\kappa = 2}$ cost. 
Proceeding with the latter, the dominant contribution to the complexity for the regulated field theory is then given by \eqref{eqn:C-N2-pen-nonper}.
The corresponding small $\tilde{\alpha}$ expansion of \eqref{eqn:C-N2-pen-nonper} has been introduced in \eqref{eqn:C-N2-pen-per}.

\subsection{Unpenalized vs penalized}
Finally, once the complexity for the regulated field in penalized geometry is given, we can compute the difference
$\Delta \D  := \pD_{\kappa = 2}  - \D_{\kappa = 2}$. From previous findings we expect the latter to be positive semidefinite.
The corresponding complexities for both geometries can be found in \eqref{eqn:C-N2-pen-nonper} and \eqref{eqn:C_kappa}. 
Using the small $\tilde{\alpha}$ expansion from \eqref{eqn:C-N2-pen-per} for the former penalized one in  \eqref{eqn:C-N2-pen-nonper}, we find that
\eqnsplit{
\Delta \D \simeq  \frac{V}{\delta^{d-1}}\lc \logb{ \frac{1}{1- \underline{\breve R} \tilde{\alpha}} } + 2 \logb{ \frac{1}{\delta \omega_0} }  \rc
\logb{ \frac{1}{1- \underline{\breve R} \tilde{\alpha}} }.
\label{eqn:DeltaD}
}
Of course, in the unpenalized limit, the expression in \eqref{eqn:DeltaD} vanishes, i.e. $\lim_{\tilde{\alpha} \rightarrow 0} \Delta \D  = 0$. 
As soon as the gates are penalized, means when $\alpha > 1$ or $\tilde \alpha > 0$, respectively, the difference in the optimal depth starts to grow, similar to the complexity increase in the two coupled oscillator case from section \ref{subsec:small-pen}.
It should be noted that the sign of the approximate $\Delta \D$ from \eqref{eqn:DeltaD} seems to depend on $\mathcal{O}(\delta \omega_0)$. 
This basically goes back to the small $\tilde \alpha$ expansion used in \eqref{eqn:DeltaD} for which we need to distinguish between the different cases.
For $\delta \omega_0 \lesssim 1$, we obviously find 
$\Delta \D \gtrsim 0$ as we would expect.
In case of $\delta \omega_0 \gtrsim 1$, we get a negative second term in the first line, but the leading positive term will ensure that $\Delta \D$ is still positive. 
However, if $\underline{\breve R} \tilde{\alpha}$ in the argument of the leading logarithm becomes too small, the logarithmic factor in the second line of course decreases either so that $\Delta \D \rightarrow 0$.
For simplifying the underlying expressions, let us assume\footnote{For instance, in cMERA, it is usually assumed that $\mathcal{O}(\delta \omega_0) \sim 1$.} $\delta \omega_0 \lesssim 1$ for the remainder of this discussion.

According to our assumptions, we may in addition drop the leading term in the first line of \eqref{eqn:DeltaD} as well as the factor 2 in front of the second term which is assumed to be dominating. By doing so, we may further estimate
\eqn{
\Delta \D \simeq \frac{V}{\delta^{d-1}} \logb{ \frac{1}{1- \underline{\breve R} \tilde{\alpha}} } \logb{ \frac{1}{\delta \omega_0} }.
\label{eqn:DeltaD2}
}
If we now carefully look at the right-hand side of \eqref{eqn:DeltaD2}, we notice that this expression is proportional to the complexity one would obtain for the $\mathcal F_{\kappa = 1}$ cost. Of course, the absolute value bars in $\D_{\kappa=1}$ can be neglected due to $\delta \omega_0 \lesssim 1$, see also discussion below. Specifically, defining $\Theta :=- \log(1- \underline{\breve R} \tilde{\alpha})$ where $0 < \Theta \ll 1$,
the relation from above can equivalently be written as
\eqn{
\pD_{\kappa = 2}  \simeq  \Theta \D_{\kappa = 1} + \D_{\kappa = 2}.
\label{eqn:rel-kappa-2-1}
} 
Again, we would end up with the unpenalized complexity if $\Theta = 0$.
Recall that the $\mathcal F_{\kappa = 1}$ cost yields the same leading order UV divergence as found via the holographic complexity proposals \cite{Jefferson:2017sdb}. Even more, this particular cost---which gives rise to the so-called Manhattan metric---leads to the same logarithmic factor predicted on basis of the CA proposal, see section \ref{subsubsec:C-holo}.

Let us here note that an interesting relation between the $\mathcal F_{2}$ and $\mathcal F_{\kappa = 1}$ costs has recently been found by exploring Nielsen's geometric approach to circuit complexity for two dimensional CFTs.\footnote{The corresponding submanifolds are associated to the underlying symmetry groups.} 
It has been shown that the difference between distinct 
metrics on the Virasoro circuits turns out to be
vanishing in the holograhic large central charge limit  \cite{Caputa:2018kdj}.

Coming back to the present case, we can also write $\mathcal D_{\kappa = 2} = \Phi   \mathcal D_{\kappa = 1}$ by simply defining $\Phi := - \log(\delta \omega_0)$. This is possible, since by assuming $\delta \omega_0 \lesssim 1$ we can write $\vert \log(\frac{1}{\delta \omega_0}) \vert = \Phi > 0$ as noted before. Following these steps and replacing $\mathcal D_{\kappa = 2}$  in \eqref{eqn:rel-kappa-2-1} appropriately then yields
\eqn{
\pD_{\kappa = 2}  \simeq  (\Theta + \Phi)   \D_{\kappa = 1}.
\label{eqn:rel-kappa-2-1-v2}
}
This relation between complexities for two different costs is interesting. It somewhat resembles the situation described above in which a relation between two different costs has been found as well.
Referring to the approximate expressions in \eqref{eqn:rel-kappa-2-1} and \eqref{eqn:rel-kappa-2-1-v2}, we may therefore ask whether weighting certain classes of gates---in particular the entangling one, since these also play an important role in holography related aspects---for a specific cost can compensate the geometry change in the space of unitaries for a different cost. 

Of course, our previous estimations rely on certain simplifications. More detailed studies concerning such aspects may be important, particularly, if we presume that $\mathcal F_{\kappa = 1}$ is the most holographic like cost which might be intrinsic to the CV and CA proposals. 
For that reason, we are therefore confronted with the question whether weighting gates in a certain way are necessary for a consistent definition of complexity in QFT. Moreover, what 
would be the corresponding holographic interpretation?

On the other hand, let us bring to mind that notions such as reference, geometry or gate set are not explicitly set in the holographic CV and CA proposals. 
Instead, they appear to be implicit in the duality. 
Therefore, identifying possible similarities between QFT and holography along these lines, might serve valuable information and shed light on the role of gate weighting.

\subsection{Divergence}
\label{subsec:div}
Resorting to the UV approximation, we have seen that there appears an additional logarithmic factor in the complexity, see expression  \eqref{eqn:C_kappa}. In the limit $\delta \rightarrow 0$, this is the contribution which diverges much faster than the prefactor $V \delta^{1-d}$, for more details see also \cite{Jefferson:2017sdb}.
In order to eliminate the additional divergence, one may, for instance, assume that the reference frequency $\omega_0$ is fixed at some UV scale with $\omega_0 \sim e^{-\sigma} / \delta$. Again, the numerical factor $\sigma$ would ensure that $\omega_0 > \omega_{\vec{k}}$. 
We should note that these estimations are of course very rough and are only meant to provide some hint about the approximate scaling behavior.
A more general approach is presented in \cite{Jefferson:2017sdb} for the unpenalized $\kappa = 1$ measure. 
Nevertheless, as a first attempt, we would like to continue in the following with the more simple estimations.\footnote{Note that, at least, referring to the $\kappa = 1$ measure, the simple estimation here still leads to the same scaling behavior as present in the more general result derived by computing the sum over all normal mode frequencies $\tilde \omega_{\vec{k}}$, cf. \cite{Jefferson:2017sdb}.} Inserting the explicit choice for $\omega_0$ from above into \eqref{eqn:C_kappa} yields
\eqn{
\mathcal C_{\text{UV}, \kappa}  \sim \sigma^\kappa  \frac{ V}{\delta^{d-1}},
\label{eqn:C_kappa-UV}
} 
so that the additional divergence is eliminated. 
As we can see, the leading order UV divergence becomes a power law. On the other hand, if we suppose to work in the IR approximation by making the following simple assumption
\eqn{
\tilde \omega_{\vec{k}} \sim  m,
\label{eqn:ir-approx}
}
then the IR contribution to the complexity takes the form 
\eqn{
\mathcal C_{\text{IR}, \kappa} \sim - \log^\kappa(m \delta)
\label{eqn:C_kappa-IR}
}
after having assumed $m\delta \ll 1$.
From the approximation above we get that the cancellation of the additional factor engenders that the IR contributions to the complexity depend on the UV cutoff scale $\delta$.

To avoid this feature, one may assume $\omega_0 \ll 1/\delta$ and identify $\omega_0$ with some IR frequency \cite{Jefferson:2017sdb}. However, in this case, the additional logarithmic factor will again contribute to the leading divergence in the limit $\delta \rightarrow 0$.

As we can see, there are certain ambiguities in the discussions above. An important point is that the scaling of the complexity clearly depends on the explicit choice for the reference frequency $\omega_0$. The dependence on any reference state parameter is obviously unsatisfying from the holographic point of view, since in the AdS/CFT dictionary this choice is not explicit (this is not so for the target state which in contrast has an holographic interpretation, see e.g. \cite{Bernamonti:2019zyy}) and we do not exactly know how to get access to it at all. 

Let us return to the UV approximation in the penalized geometry for the $\mathcal F_{\kappa = 2}$ cost from \eqref{eqn:C-N2-pen-nonper}. Assigning a higher cost to the entangling gates introduces an additional degree of freedom which is the weighting factor $\alpha$. 
Contrary to the frequency $\omega_0$, whose explicit value plays a substantial role in the unpenalized case, the factor $\alpha$ does not characterize the reference state.
It specifies the metric in the manifold of unitaries. 
Hence, instead of fixing the reference frequency and thus the reference state in a certain way, we may simply eliminate the additional logarithmic contribution by tuning $\alpha$ correspondingly. 

To derive the explicit form of $\alpha$, one may solve a similar equation as before, i.e.
\eqn{
\bdelta_* \omega_0 = e^{-\nu}
\label{eqn:rel-eqn-nu}
}
where $\nu$ is again taken to be some overall numerical factor. The leading contribution to the complexity would then read\footnote{As in the unpenalized case, although this rough estimation is in the first instance sufficient to deduce the characteristic leading order scaling, as well as the subleading IR contribution, a more detailed investigation will be interesting.}
\eqn{
\mathcal C_{\text{UV}, \kappa = 2}  \sim  \nu^2  \frac{V}{\delta^{d-1}},
\label{eqn:C_2-UV-alpha}
}
as in \eqref{eqn:C_kappa-UV}, i.e. without the additional logarithmic divergence dominating in the limit $\delta \rightarrow 0$. 
Thus, avoiding any specific choice for $\omega_0$, 
a simple power law for the leading divergence multiplied by an arbitrary factor
can be realized via choosing $\alpha$ to be the solution of  \eqref{eqn:rel-eqn-nu}.
In the mentioned UV approach with $m \ll 1/\delta$, it can be shown that
the weighting factor is determined by
\eqn{
\alpha \tan^{-1}(\sqrt{\alpha^2 - 1}) \simeq \frac{  \log^2  \lc \frac{4}{e^{-2 \nu}} +  \frac{m^2}{\omega_0^2}  \rc -  \log^2 \lc \frac{1}{\delta^2 \omega_0^2} \rc }{ 4  \log \lc \frac{ 1 }{\delta^2 m^2} \rc }.
\label{eqn:alpha-approx}
}
Of course, in the unpenalized case, i.e. $\alpha = 1$, in other words, if we suppose $\omega_0$ to be the only tunable parameter, we simply get $\nu = \sigma$ according to \eqref{eqn:alpha-approx}.

Similarly, we may focus on the IR contribution in the penalized case. In the latter case we may derive an analogous expression as in \eqref{eqn:C_kappa-IR}, where we need to take into account the modified parameter $\bdelta_*$.
Apart from the IR contribution, according to the generalization derived for the unpenalized $\mathcal F_{\kappa=1}$ cost by summing over the normal mode frequencies \cite{Jefferson:2017sdb},
we may also expect some subleading UV contributions to the complexity which would roughly scale as
\eqn{
\mathcal C_{\text{UV-sub}, \kappa} \sim (m_* \delta)^{2}
\label{eqn:C_kappa-UVsub-alpha}
}
in the small $\tilde{\alpha}$ limit. Let us emphasize that, just formally, $m_* \equiv m \lc 1- \underline{\breve R} \tilde{\alpha}  + \mathcal{O}(\tilde{\alpha}^2) \rc$ may be seen as some effective mass which reduces to $m$ when the entangling gates are not penalized, i.e. if we take the limit $\tilde{\alpha} \rightarrow 0$.

\subsection{Comparison with $\mathcal C_\text{A}$}
We can also compare the result in \eqref{eqn:C-N2-pen-nonper} with the prediction \eqref{eqn:C-CA} based on the CA proposal where an extra logarithmic factor arises due to the asymptotic joint contributions in the WDW action \cite{Carmi:2016wjl}. Recall that in the unpenalized case, the QFT prediction \eqref{eqn:C_kappa} follows if the dimensionless coefficient $\Lambda$ is set equal to $\omega_0 L_\text{AdS}$. This choice may be motivated by the fact that $\mathcal C_\text{A}$ should be independent of the AdS curvature scale $L_\text{AdS}$ when the complexity is defined in the CFT.

In the penalized case, the comparison between the logarithmic factors becomes even more interesting. As before, we may identify the CFT cutoff scale in \eqref{eqn:C-CA} with the scale parameter $\delta$ appearing in \eqref{eqn:C-N2-pen-nonper}. Then, taking, for instance, the small $\tilde{\alpha}$ expansion from \eqref{eqn:C-N2-pen-per}, we may relate both contributions to each other. The cancellation of the bulk curvature scale $L_\text{AdS}$ may be realized if we set
\eqn{
\Lambda = (1- \underline{\breve R} \tilde{\alpha}) \omega_0 L_\text{AdS}.
\label{eqn:Lambda-elim}
}
Inserting $\Lambda$ from \eqref{eqn:Lambda-elim} into the general expression \eqref{eqn:C-CA} then yields
\eqn{
\mathcal C_\text{A} \sim \frac{V}{\delta^{d-1}} \log \lc \frac{1}{\delta \omega_0 (1 - \underline{\breve R} \tilde{\alpha}) } \rc.
\label{eqn:C-A-small}
}
In contrast to \eqref{eqn:C-CA-replaced}, we end up with an additional contribution in the denominator which is proportional to $\breve R \tilde{\alpha}$ and thus depends on the weighting factor $\alpha$. Of course, the expression \eqref{eqn:C-A-small} still depends on the unspecified reference frequency $\omega_0$. At least when gate penalties are not implemented, we have encountered that this turns out to be a general feature of the notion of circuit complexity in QFT. 
Recall that in the unpenalized situation, eliminating the additional logarithmic contribution required fixing $\omega_0$ at an appropriate scale. 

However, introducing gate penalties, we have shown that tuning the weighting factor appropriately can eliminate the logarithmic contribution as well.
Now, the logarithmic contribution in $\mathcal C_\text{A}$ from \eqref{eqn:C-A-small} can be eliminated in the same way. In order to do so, we may for instance demand \eqref{eqn:rel-eqn-nu} to be fulfilled. In the present perturbative approximation (with respect to $\tilde{\alpha}$), we would need to solve
\eqn{
\delta \omega_0 (1- \underline{\breve R} \tilde{\alpha}) = e^{-\nu}
} 
to find the required $\alpha$ which eliminates the additional logarithmic factor. The latter would then yield the leading order contribution
\eqn{
\mathcal C_\text{A} \sim \nu \frac{V}{\delta^{d-1}}.
}

As mentioned, the dimensionless coefficient $\Lambda$ is related to the null normals on the boundary of the WDW patch in the bulk. To eliminate the curvature scale of the bulk spacetime $L_\text{AdS}$, we have set  $\Lambda$ as in \eqref{eqn:Lambda-elim}.
Referring to such a relation between $\Lambda$ and $\alpha$ (note that $\tilde{\alpha}^2 = \alpha -1$), we may ask how weighting a certain class of gates would be translated via holography. 
Of course, the present studies are clearly too premature to speculate towards this direction, but the found relations from the present and the previous sections may motivate further investigations.

Note that our findings have been obtained for a regulated free field. It is clear that this is far from dealing with strongly coupled theories with a large number of degrees of freedom with existent holographic duals. However, choosing the cost appropriately, the resulting similarities with holographic complexity proposals for the leading order UV divergence may provide useful insights. Namely, as noted above, such an accordance may suggest which cost function is intrinsic to the holographic complexity conjectures.
Indeed, this was actually the main reason behind utilizing the $\mathcal F_{\kappa}$ costs.

Despite the mentioned concordance, it will actually be necessary to extend the notion of complexity for strongly coupled theories with holographic duals, or at least, for some suitably excited states going beyond the free case.
As mentioned, in the context of CFTs, interesting progress has been done in \cite{Caputa:2018kdj}, also see \cite{Belin:2018bpg}. Note also the recent studies in \cite{Caputa:2017urj,Caputa:2017yrh,Czech:2017ryf} introducing a notion of complexity which might shed light on this issue as well.
Beyond these, circuit complexity has recently been studied for the TFD state in a free scalar field theory \cite{Chapman:2018hou}, also see \cite{Yang:2017nfn} for similar considerations. The TFD state formed by entangling two copies of a CFT is especially important in holography, since, as already pointed out below figure \ref{fig:cv-ca}, it is dual to an eternal BH in AdS \cite{Maldacena:2001kr}.

Thus, it would be interesting to investigate the role of gate penalties in more holographic like setups in order to work out their possible holographic interpretation. On the other hand, as stressed before, such systems would particularly be interesting for making direct comparisons with recent studies arguing that holographic complexity might be nonlocal \cite{Fu:2018kcp}. 

\subsection{Distance dependent penalties}
To find out what type of conditions are generically needed to be fulfilled by a field theory so that it has a bulk dual is important for better understanding the basic mechanisms behind holography. In previous studies, interesting conditions have been worked out by analyzing the corresponding boundary CFT, see e.g. \cite{Heemskerk:2009pn,ElShowk:2011ag}.

More recently, it has been discussed that investigting the complexity growth for certain type of gauge theories may also shed light on such necessity conditions \cite{Hashimoto:2017fga}. The motivation here is the conjectured criterion for the existence of a dual BH, namely, an upper bound in complexity with the scaling behavior $\mathcal{C}_\text{max} \sim e^S$ where the exponent $S$ denotes the corresponding entropy. Indeed, it has been shown that for both the classical as well as quantum cases, the complexity first grows and saturates afterwards at the maximum $\mathcal{C}_\text{max}$ being in accordance with the second law of complexity \cite{Brown:2016wib,Brown:2017jil}. 
More precisely, it has been found that the speed of growth increases when the time evolution is more nonlocal. It has been found that only the maximally nonlocal gauge theories satisfy the mentioned scaling behavior from above and thus possess the possibility of a gravity dual. 
Accordingly, the more nonlocal the interaction is the higher is the possibility for the existence of an holographic dual of the underlying theory.

On the other hand, as already mentioned, it has been discussed that complexity holographically defined via the CV and CA proposals may be nonlocal as well, but in a slightly different manner.
To be more specific, it has been argued that any gate set for a CFT defining holographic complexity should necessarily contain bilocal gates acting at arbitraryly separated points \cite{Fu:2018kcp}. 
Viewing the both conjectures as possible entries in the holographic dictionary, the presence of such bilocal gates constructing the corresponding field theory state may be taken as another necessity condition, similar to the one described above.

In the present paper, we use such kind of bilocal gates as well. However, we implement gate penalties for the entangling gates where those are uniformly weighted, thus, distinguishing between the different gate classes. It turns out that as long as we insert such distance independent penalties, we obtain a similar divergence structure in the complexity as seen in holographic computations, see \eqref{eqn:C-N2-pen-nonper}. However, following the discussion in section \ref{sec:pen-qft}, this kind of agreement might not appear if certain gates are penalized according to the distance of points they act on. 
Even more, one may for instance think of entirely preventing such gates from participating in the construction of the circuit.
In either cases, the characteristic power law scaling of the leading order UV divergence $\mathcal{C}_{\text{UV},\kappa} \sim V\delta^{1-d}$, but of course the additional logarithmic factor as well, might be modified or even be rescinded completely.
Introducing locality in the way as described above may therefore lead to drastic differences between complexity in field theory and the holographic proposals. Taking into account that certain costs may be intrinsic to the holographic proposals due to the found similarities, such expectations may indeed favor the existence of bilocal gates in the gate set for holographic states as argued in \cite{Fu:2018kcp}. 
Apart from that, the existence of bilocal gates in recent studies of circuit complexity for interacting \cite{Bhattacharyya:2018bbv} and time dependent nonlocal states \cite{Ali:2018fcz} may explain the found behavior which is in line with the holographic predictions.
It remains interesting to find out whether and how implementing distance dependent gate penalties would affect such agreements.

\section{Conclusion}
\label{sec:conc}
We have investigated the effects of weighting certain classes of quantum gates constructing a quantum circuit for states in QFT.
Our studies are based on the geometric approach to circuit complexity proposed by Nielsen and collaborators.
Introducing such gate penalties may be motivated from various perspectives. For instance, they may incorporate the notion of locality in complexity when the latter is taken to be a physical attribute of a QFT. 
On the other hand, 
they may have important implications in the holographic dictionary. 
Closely related, also an application for tensor networks which are believed to provide a representation of a time slice of the AdS bulk space may motivate considering a penalized geometry in the space of unitaries. 

In the present work, we have not introduced distance dependent weighting factors. Instead, we have worked with entangling gates which have been uniformly weighted. Of course, this does not introduce the notion of locality
as described above. Nevertheless, an implementation of such kind has enabled us obtaining interesting insights which may motivate further investigations, in particular, by examining similar ideas for strongly coupled theories with holographic duals.
Our results have generalized earlier findings dealing with the case of a pair of two coupled harmonic oscillators.
More specifically, we have made the extension to the case of a regulated free field theory placed on the lattice for which the optimal circuit acts in form of a representation of $GL(N,\mathbb{R})$.

Comparing both the penalized and the unpenalized results with each other, we have seen that assigning a higher cost to the entangling gates gives rise to a substantial increase in complexity. 
Furthermore, using the complexities for the $\mathcal F_{\kappa = 2}$ cost function, we have found an interesting relation between these and the complexity for the unpenalized $\mathcal F_{\kappa=1}$ cost which is known to give rise to the Manhattan metric. From earlier studies we know that $\mathcal F_{\kappa=1}$ behaves as the most holographic like cost, since it yields the same leading order UV divergence as obtained via the holographic complexity proposals, including the additional logarithmic factor as appearing in the CA prediction. Referring to earlier results, the found relations have led us to speculate whether the change in the manifold of unitaries determined by a specific cost function may be compensated by introducing a penalty tensor for a different cost.

In addition, we have exhibited how the gate weighting modifies the leading order UV divergence in the complexity. 
In contrast to the unpenalized geometry, penalizing the entangling gates has introduced an additional degree of freedom, the latter corresponding to the mentioned weighting factor. 
We have shown that appropriately tuning this factor can eliminate the mentioned logarithmic contribution. Importantly, this procedure turned out to be independent from the choice for the reference frequency. Recall that in the penalized case, the latter is the only tunable control parameter. The dependence on any state parameter of this type is not explicit in the holographic proposals. Hence, it is of particular interest that the divergence can be modified by tweaking the weighting factor which specifies the metric in the space of unitaries instead of characterizing the reference state. In earlier studies it has been discussed that the geometric structure in the mentioned space may encode important features of the holographic proposals. The main reason for such ideas lies in the found similarities in the divergence structure for some specific costs such as the $\mathcal F_{\kappa}$ one. Indeed, these similarities may suggest which cost is intrinsic to the holographic complexity conjectures. Hence, referring to our findings, the insertion of gate penalties in the space of unitaries makes such connections become even more interesting.

We have also compared our field theory results with the mentioned holographic proposals where we have given particular consideration to the predictions based on the CA conjecture. By comparing both predictions with each other, we have related the weighting factor to quantities which are connected to the WDW patch in the bulk AdS spacetime.
In view of these findings, we have finally commented on certain speculative expectations concerning the role of gate penalties in defining complexity for states in QFT and their possible implications in the holographic context.

\section{Acknowledgements}
I would like to thank Arpan Bhattacharyya, Pawel Caputa and Tadashi Takayanagi for interesting conversations and helpful comments on this work. 
I also thank Mario Flory and Joan Simon for earlier conversations and Tadashi Takayanagi for the generous hospitality at the Yukawa Institute for Theoretical Physics at Kyoto University.

\appendix

\bibliography{article_bib}

\end{document}